\documentclass[a4paper,fleqn,usenatbib]{mnras}
\usepackage[T1]{fontenc}
\usepackage{ae,aecompl}
\usepackage{hyperref}

\usepackage{graphicx}	% Including figure files
\usepackage{amsmath}	% Advanced maths commands
\usepackage{amssymb}	% Extra maths symbols

\usepackage{newtxtext,newtxmath}
\usepackage{subfig}
\usepackage{xcolor}
\usepackage{multirow}
\usepackage{float}
\usepackage[symbol]{footmisc}

\title[ACF and PDM for period searching in AGN]{Detection of periodic signals in AGN red noise light curves: Empirical tests on the Auto Correlation Function and Phase Dispersion Minimization}

\author[Krishnan et al.]{
S. Krishnan,$^{1}$\thanks{E-mail: saikruba@camk.edu.pl}
A.G. Markowitz,$^{1,2}$
A. Schwarzenberg-Czerny,$^{1}$
M.J. Middleton $^{3}$
\\
$^{1}$Nicolaus Copernicus Astronomical Center, Polish Academy of Sciences, Bartycka 18, 00--716 Warsaw, Poland\\
$^{2}$Center for Astrophysics and Space Sciences, Univ.\ of California, San Diego, MC 0424, La Jolla, CA 92093-0424, USA\\
$^{3}$Department of Physics and Astronomy, University of Southampton, Southampton SO17 1BJ, UK
}

\pubyear{2020}

\begin{document}
%\iffalse
\label{firstpage}
\pagerange{\pageref{firstpage}--\pageref{lastpage}}
\maketitle

\begin{abstract}
Active galactic nucleus (AGN) emission is dominated by stochastic, aperiodic variability which overwhelms any periodic/quasi-periodic signal (QPO) if one is present. The Auto Correlation Function (ACF) and Phase Dispersion Minimization (PDM) techniques have been used previously to claim detections of QPOs in AGN light curves. In this paper we perform Monte Carlo simulations to empirically test QPO detection feasibility in the presence of red noise. Given the community's access to large databases of monitoring light curves via large-area monitoring programmes, our goal is to provide guidance to those searching for QPOs via data trawls. We simulate evenly-sampled pure red noise light curves to estimate false alarm probabilities; false positives in both tools tend to occur towards timescales longer than (very roughly) one-third of the light curve duration. We simulate QPOs mixed with pure red noise and determine the true-positive detection sensitivity; in both tools, it depends strongly on the relative strength of the QPO against the red noise and on the steepness of the red noise PSD slope. We find that extremely large values of peak QPO power relative to red noise (typically $\sim 10^{4-5}$) are needed for a 99.7 per cent true-positive detection rate. Given that the true-positive detections using the ACF or PDM are generally rare to obtain, we conclude that period searches based on the ACF or PDM must be treated with extreme caution when the data quality is not good. We consider the feasibility of QPO detection in the context of highly-inclined, periodically self-lensing supermassive black hole binaries.

\end{abstract}


\begin{keywords}
methods: statistical -- galaxies: active 
\end{keywords}

%%%%%%%%%%%%%%%%%%%%%%%%%%%%%%%%%%%%%%%%%%%%%%%%%%

%%%%%%%%%%%%%%%%% BODY OF PAPER %%%%%%%%%%%%%%%%%%

\section{Introduction}
Active galactic nuclei (AGN), driven by matter accreting onto supermassive black holes (SMBHs), are among the most powerful and steady sources of luminosity in the Universe. AGNs are luminous across the electromagnetic spectrum (EM) --- bolometric luminosities can span typically $10^{40}$ -- $10^{47}$ erg s$^{-1}$ --- particularly in the optical/UV regime. In many cases, emission in the radio and gamma-ray bands due to collimated jets is also observed. Since their discovery, AGNs have featured in a range of both theoretical and observational insights: Correlations between SMBH masses and both the stellar velocity dispersions and luminosities of host galaxy bulges suggest co-evolution of SMBHs and their host galaxies (\citealt{1998AJ....115.2285M}; \citealt{2000ApJ...539L...9F}). AGNs likely provide galaxy-scale radiative and mechanical feedback (e.g. \citealt{2007ARA&A..45..117M}). Additionally, AGNs could serve as excellent testing grounds for general relativity through the investigation of dynamical accretion flows, and supermassive binary black hole systems might potentially reveal a wealth of information via multi-messenger signals.

AGN continuum emission can be strongly variable on timescales from ks to years; study of this continuum variability can elucidate characteristic timescales that provide insight into accretion flow or jet physics. Variability mechanisms are likely linked to system parameters such as black hole mass $M_{\rm BH}$, luminosity, accretion rate relative to Eddington, etc. For example, the characteristic timescales measured in broadband X-ray power spectral density (PSD) functions (``breaks'' in the continuum power-law slope; \citealt{1999ApJ...514..682E}; \citealt{2002MNRAS.332..231U}; \citealt{2003ApJ...593...96M}) scale with both $M_{\rm BH}$ and luminosity, as empirically quantified by \citet{2006Natur.444..730M}. The extrapolation of this relation to stellar-mass black hole X-ray Binaries (BHXRBs) --- along with X-ray/radio luminosity relations (\citealt{1999ApJ...514..682E}) --- supports the notion of identical accretion mechanisms in both classes of objects, and leads to the interesting possibility that variablity components present in one class of object may be present in the other. The X-ray PSDs of several actively-accreting BHXRBs also reveal quasi-periodic oscillations (QPOs; e.g. \citealt{1999ApJ...526L..33W}; \citealt{2004A&A...426..587C}; \citealt{2015MNRAS.447.2059M}). The so-called low-frequency QPOs typically observed near frequencies $f$ $\sim$ 1--30 Hz, evolve in frequency as source luminosity and inner disk size evolve and may be associated with Lense-Thirring precession in the inner disk (e.g. \citealt{2012MNRAS.419.2369I}).  High-frequency QPOs usually occur at 40 -- 450 Hz; the frequencies scale as the inverse of $M_{\rm BH}$, and thus may be an imprint of $M_{\rm BH}$ and spin (\citealt{1999ApJ...522..397R}; \citealt{2001A&A...374L..19A}). 

The efforts to locate periodic signals (either strictly- or quasi-periodic) in both Seyfert AGN and blazars have involved studies using light curves spanning the EM spectrum. Some interpretations posit a ``hot spot'' in the innermost accretion disk, yielding inferred constraints on the size of the inner disk or on $M_{\rm BH}$ (e.g. \citealt{2009ApJ...690..216G}). Some periodicities reported in blazars are interpreted as due to jets' precessing in and out along the line of sight (e.g. \citealt{1999A&A...347...30V}; \citealt{10.1086/648433}; \citealt{2016AJ....151...54S}) including modulation associated with the blazar being part of a gravitationally-bound SMBH binary system (e.g. \citealt{1996ApJ...460..207L}; \citealt{2006ApJ...643L...9V}). Additional recent claimed periods for quasars have also been interpreted as support for binary SMBH systems; luminosity variations are ascribed to modulations in the mass accretion rate caused by the binary's orbital motion (e.g. \citealt{Liu_2015}; \citealt{2015Natur.518...74G}; \citealt{2016MNRAS.463.2145C}).

However, \textit{statistically robust} detections of strictly-periodic oscillations (SPOs) or QPOs remain a challenge. Limited data quality for AGN can pose a challenge. For example, searches for QPOs in the broadband X-ray PSDs or periodograms\footnote{The periodogram is an estimator of the PSD.} of Seyferts is hampered by poor frequency resolution, particularly in comparison to the higher-quality X-ray timing data for BHXRBs in which QPOs are typically detected \citep{2005MNRAS.362..235V}. Hence, so far there have been only a few robust X-ray-based claims, such as the QPO observed in RE~J1034+396, which persists at a timescale of $\sim1$ hr across observations spanning years (\citealt{2008Natur.455..369G}; \citealt{2014MNRAS.445L..16A}) a $\sim2$ hr QPO detected in MS 2254.9$-$3712 (\citealt{2015MNRAS.449..467A}) and a $\sim$ 24 minute QPO in IRAS 13224$-$3809 (\citealt{2019MNRAS.482.2088A}); for additional detections see \citet{2021MNRAS.501.5478A}. However, the behavior of the PSD --- e.g. distribution and scatter of periodogram points; biases --- under a variety of noise processes (white, red, etc.) and in astrophysical contexts is now well-understood (e.g. \citealt{1929RSPSA.125...54F}; \citealt{1983ApJ...266..160L}; \citealt{1993MNRAS.261..612P}). Particularly when data quality is high, meaning data sampling is relatively continuous (few major gaps) and close to even spacing, the periodogram or PSD is straightforward to use for detections of QPOs (e.g. \citealt{2005A&A...431..391V}). Otherwise alternate statistical methods are generally employed for the detection of periodic/QPO signals in AGN with sparsely sampled data points, such as the Auto-Correlation Function (ACF), Phase Dispersion Minimization (PDM), wavelet analysis, sinusoidal fitting, etc. Generally, QPOs claimed in AGN using these alternate methods are non-repeatable in additional observations, are based on improper usage of statistical tools, such as simply identifying the highest-amplitude points in a periodogram as outliers (e.g. \citealt{1988AJ.....95..374W}; \citealt{1995PASP..107..863S}; \citealt{2013MNRAS.434.3122P}; \citealt{2015ApJ...813L..41A}), or improper calibration of the ``false alarm probability.''  In particular, the presence of stochastic, aperiodic ``red noise'' is seen at all wavebands and dominates the AGN emission and  which tends to bury any possible periodic/quasi-periodic signal, if any present. In addition, pure stochastic red noise processes (no QPO present) are seen to spuriously mimic few-cycle sinusoid-like periods  (\citealt{2010ApJ...708..927K}; \citealt{2016MNRAS.461.3145V}), which can be misinterpreted as an intrinsic periodic signal. Hence, it is crucial to account for the amount and form of the red noise which can impact the calculation of statistical significances of detection of periods and calibration of false alarm probability while using any statistical tool. Many claims of AGN periodicities in the literature simply made no attempt to account for this red noise ``background.'' In addition, many claims of AGN periodicities in the literature are few-cycle (e.g. \citealt{2015ApJ...813L..41A}; \citealt{2016ApJS..225...29B}). In either of these scenarios, it is likely the claimed periodicity does not exist, and is merely an artifact of red noise. Finally, the accumulated publications of periods in AGN do not yet exhibit any obvious trend between timescale and system parameters such as $M_{\rm BH}$ or accretion rate relative to Eddington.  In contrast, the continuum break timescales in X-ray PSDs and the characteristic frequencies of low-frequency QPOs and broad Lorentzian components in BHXRBs are observed to robustly depend on $M_{\rm BH}$ and/or accretion rate.

We are in the era of ``Big Data'', with the ground-based observing programmes such as PanSTARRS, the Palomar Transient Factory, and LOw Frequency ARray (LOFAR) and near-future programmes such as the Vera C.\ Rubin Large Synoptic Survey Telescope (LSST), Zwicky Transient Factory (ZTF), and Square Kilometer Array (SKA) now monitor or will monitor large fractions of the sky. The resulting light curve databases are likely to enable period searches over $10^3$ to $10^6$ AGN simultaneously, and examples of period searches resulting from such database trawls already exist (e.g. \citealt{2015Natur.518...74G}; \citealt{2016MNRAS.463.2145C}). If false alarm probabilities for given statistical tests are not properly calibrated, then tests run on a statistically large sample can potentially yield false detection rates. Our goal is to provide guidance for AGN QPO searches and publications, so in this work, we formulate guidelines on the proper use of two commonly-used statistical methods --- the Auto-Correlation Function (ACF) and Phase Dispersion Minimization (PDM) --- in red noise-dominated AGN light curves. We test their efficacy in robustly distinguishing between a pure stochastic red noise process (no QPOs) and a mixture of red noise and a high-quality QPO signal.

This paper is arranged as follows: In $\S$2, we review our methodology for light curve simulations. In $\S$3 and $\S$4, we present results for the ACF and PDM, respectively, for both pure red-noise processes and mixtures of red noise plus QPOs, for an evenly-sampled light curve. In $\S$5, we explore the effects of gaps and irregular sampling patterns in the light curve while using the PDM. We review results in light of interpretation and application to physical systems in $\S$6. A summary of our main results is presented in $\S$7.

\section{Methodology: light curve simulation and input PSD models}
Our method is to assume certain forms for the underlying PSD (continuum shape, presence/lack of QPOs), simulate light curves, run the ACF or PDM on these light curves, and accumulate statistics on whether QPOs are detected. With regards to PSD model shape, we can consider that  radio-quiet AGN have highly similar accretion processes to those occurring in BHXRBs, and that variability processes in both object classes are likely to be identical. In the X-ray PSDs of BHXRBs, broadband continuum shapes and narrow-band QPOs are frequently modeled with cut-off powerlaw models and/or broad or narrow Lorentzian components (e.g. \citealt{2000MNRAS.318..361N}; \citealt{2002ApJ...572..392B}; \citealt{2003A&A...407.1039P}; \citealt{2005A&A...438..999A}; \citealt{2014A&A...565A...1G}; \citealt{2015MNRAS.454.2360D}). However, in both Seyfert (disk emission dominates) and blazar (jet emission dominates) AGN, data quality for PSD measurement at all wavelengths is generally more poor compared to those for BHXRB X-ray PSDs. For Seyferts, modeling of broadband X-ray PSDs generally suffices to use unbroken power laws or broken or slowly-bending power laws; an exception is the X-ray PSD of the Seyfert 1 Ark$\sim$564, which is fitted with two broad Lorentzians (\citealt{2007MNRAS.382..985M}). For blazars, broadband PSDs at multiple wavelengths have been measured to roughly follow power-laws (e.g. \citealt{2017ApJ...837..127G}; \citealt{2018ApJ...863..175G}; \citealt{2019Galax...7...73G}) or bending power-laws \citep{2018ApJ...859L..21C}. In this paper, we perform all analyses assuming that the red noise PSD continuum can be modeled as a simple power law or occasionally as a broken power law, as is typical for most AGN data, and/or as is suitable for relatively limited dynamic ranges in temporal frequency. However, we advise readers to explore specific PSD continuum shapes and perform their own Monte Carlo simulations, if needed. We simulate time series similar to AGN light curves by using the algorithm developed by \citet{1995A&A...300..707T}. In this method, non-deterministic linear time series are produced by randomizing both the phase and amplitude of the Fourier transform of the input data of an arbitrarily chosen PSD shape. The method of \citet{1995A&A...300..707T} produces light curves adhering to a Gaussian flux distribution; readers are referred to \citet{2013MNRAS.433..907E} for simulating light curve corresponding to arbitrary (non-Gaussian) flux distributions. For all input PSD models, we adopt the ``RMS$^2$/Hz'' normalization of \citet{1991ApJ...383..784M} and \citet{1997scma.conf..321V}.

For both the ACF and PDM, we first consider an ``ideal case'' of a QPO only, in the absence of any broadband noise, to indicate the signatures that one can pursue when searching for a QPO.  We thus consider as our model input a narrow Lorentzian component whose PSD is given by
\begin{equation}
P_{\rm Lor} = \frac{2R^2 Q f_L   }{\pi (f_L^2 +  4 Q^2 (f-f_L)^2)}
\end{equation} 
where $f_{\rm L}$ is the centroid frequency, and $R$ is the fractional RMS amplitude (absolute RMS/mean). $Q$ is the ``quality factor'' for quantifying the width of the Lorentzian, where a relatively higher value of $Q$ means a more narrow-peaked Lorentzian ($Q \sim f_{\rm L}$/FWHM). A strictly-sinusoidal, perfectly coherent oscillation has $Q = \infty$.

We also simulate light curves corresponding to pure red noise only, as modeled with a simple unbroken power-law PSD: 
\begin{equation}
P_{\rm PL} = A$  $(f/(10^{-6} {\rm Hz}))^{-\beta}
\end{equation}

$A$ is the amplitude in units of Hz$^{-1}$ at an arbitrary frequency of $10^{-6}$ Hz; in the X-ray PSDs of Seyferts, power is typically of order $10^{3-4}$ Hz$^{-1}$; we adopt $A$ = 1.5 $\times10^{4}$ Hz$^{-1}$, $\beta$ is the red noise power-law slope, where more positive values correspond to steeper slopes. We search for signatures in the ACF or PDM that could be mistaken for a QPO.  That is, by testing a range of PSD power-law slopes $\beta$, such simulations enable us to explore the false alarm probability associated with Type I errors (false positive detection of a QPO using ACF/PDM) for a range of pure stochastic processes.

Finally, we consider mixtures of red noise and QPO processes: We model variability whose PSD is described by the sum of a broadband power-law continuum plus a narrow Lorentzian for a range of power-law slopes and Lorentzian RMS strength. 

Many signals claimed in literature e.g., for claims of QPO from SMBH binaries lack estimates of the quality factor Q and are frequently assumed to be strict periods for simplicity. For example, in self-lensing SMBH binaries, the variations in the light curves could be strictly periodic though not with consistent wave amplitude due to variations in the accretion disk continuum luminosity. However, possible effects such as additional variability modes associated with the binary orbital motional/tidal interactions or the effects of periodic streams of matter impacting the disk may contribute to quasi-periodic components in the observed emission. In the case of precessing jets of blazars, some papers (e.g. \citealt{2000A&A...355..915A}; \citealt{2018MNRAS.478.3199B}) describes that the precession yields strictly periodic behavior in the resultant light curves due to periodic beaming assuming a constant input (pre-beaming) flux. But it is easy to envision how non-sinusoidal behavior can occur: the flux that gets amplified could depend on how many individual knots are being ejected at certain phases of precession, and there can be divergence in the fluxes and ballistic behavior of individual jet knots.

Since it is not possible to explore all the types of QPO signals having different widths, for our study we have chosen a relatively high quality factor, $Q=30$, to model something close to high quality periodic signals. We refer below to $P_{\rm rat}$, defined as the ratio of the $P_{\rm Lor}$($f_{\rm L}$), the power of the Lorentzian at $f_{\rm L}$, to $P_{\rm PL}$($f_{\rm L}$), the power in the power-law continuum at that frequency:
\begin{equation}
P_{\rm rat} = P_{\rm Lor}/P_{\rm PL} = \frac{2R^2Q}{\pi f_L A \left( \frac{f_L}{1.0\times10^{-6}} \right)^{-\beta}}
\end{equation}
Below, we test QPO detection thresholds for values of log($P_{\rm rat}$) spanning $-$1 to 5. 

It is impossible to consider all possible light curve sampling patterns. For our initial tests in $\S${3}, $\S${4} and $\S${5}, we assume what we refer to as our ``baseline'' sampling: evenly-sampled light curves with a duration of 250 days and having one point per day, corresponding to one representative ground-based optical observing season between 115-day yearly sun gaps. In lieu of testing potential QPOs at all frequencies between 1/(250 days) and the Nyquest frequency of 1/(2 days), we choose three representative test frequencies $f_{\rm L}$:  2.0, 8.0, and 32.0 $\times10^{-7}$ Hz, corresponding to timescales of 57.9, 14.5, and 3.6 days, and spanning  4.3, 17.2, and 69.0 cycles, respectively; we refer to them low-, medium- and high-frequency (LF, MF, and HF) QPOs (not to be confused with the LF and HF QPOs routinely identified in the X-ray PSDs of BHXRBs near typically 0.1 -- 30 Hz and 40--450 Hz, respectively). 

Finally, we neglect the effect of Poisson noise, the measurement uncertainty associated with photon counting. In the power spectrum, this noise term is a constant level of power impacting all frequencies.  In the presence of red noise, its presence likely impacts only the highest frequencies, but the exact range of course depends on both red noise slope and the level of power due to Poisson noise. As this power is another source of continuum noise, the effect would be to reduce the true-positive detection likelihood for a given value of $P_{\rm rat}$, which is defined relative to the level of red noise at that frequency. Our results can thus be considered as a "best case" scenario for true-positive detections in this regard. Given that each observation has a different level of Poisson noise, different red noise PSD shape, etc., testing all such scenarios is not feasible for the current paper, but we \textit{strongly} encourge readers to perform their own simulations taking into account the true PSD continuum shape as it is affected by the power due to Poisson noise.

\textit{As we are testing only a few ``representative'' sampling patterns, we leave it to users to run their own Monte Carlo simulations (MCS) of both pure red noise processes (for a range of continuum red noise model shapes, e.g., testing unbroken and broken/bending power-law models as necessary) and mixtures of red noise+QPOs, following our paper as a guide, and testing their own sampling patterns, candidiate QPO frequencies, and signal-to-noise values. More specifically, users are advised to test suitable ranges of red noise shapes (both in power-law continuum slopes and normalizations), QPO test frequencies, and values  of log($P_{\rm rat}$).}

\section{ACF ANALYSIS FOR PERIOD SEARCHING}
%Introduction part of ACF
The Cross Correlation Function is used for estimating the correlation coefficient of two light curves, which gives an estimate of the measure of the strength of the correlation as a function of shift in time. A positive peak implies a correlation while a negative peak implies an anti-correlation. When we use the same data set to determine the correlation function it is called an Auto-Correlation Function. There is a peak at zero lag and successive peaks at different time lag indicates the time at which the signal is correlated with itself and can be interpretated as the period of the underlying periodic/quasi-periodic signal if it exists. For a pure sinusoidal input signal, the resulting ACF is a simple cosine, with the first peak and the higher-order harmonics indicating the period. However, the two light curves are required to be evenly sampled for computing the CCF. The Discrete Correlation Function method introduced by \citet{1988ApJ...333..646E} is suitable for light curves with uneven sampling. The unbinned discrete correlation function is determined using:
 
\begin{equation}
UDCF_{ij} = \frac{(a_i-\bar{a})(b_j-\bar{b})}{\sigma_{a}\sigma_{b}} 
\end{equation}

where $a_i$ and $b_j$, are the two data sets for which we want to calculate the correlation, $\bar{a}$ and $\bar{b}$ are the mean values of the respective data set, $\sigma_{a}$ and $\sigma_{b}$ are their respective standard deviations. The DCF is determined by binning the above equation in time for each time lag $\tau$:

\begin{equation}
DCF(\tau) = \frac{1}{M}\sum UDCF_{ij}(\tau)
\end{equation}
where $M$ is the total number of pairs.

The $z$-discrete correlation function (ZDCF) \citep{2013arXiv1302.1508A} is a new method to compute the CCF when the data sets are very sparsely and irregularly sampled. It uses equal population binning and Fisher's $z$-transform to compute a more accurate estimate of the CCF: the resulting bias is more moderate and tends to zero as the number of points increases. We use the ZDCF for all the tests below.

The Interpolated Correlated Function and its associated bootstrap error (ICF; \citealt{1994PASP..106..879W}; \citealt{1998PASP..110..660P}) is also used for unevenly-sampled data. It linearly interpolates between data points and resamples to achieve an evenly-sampled light curve. The risk, as described by \citet{1994PASP..106..879W}, is that results can be misleading if the interpolation does not correctly approximate the intrinsic red-noise behavior. Given that we test a wide range of red-noise power-law slopes, including slopes as steep as $\beta=3$, we adhere to using the ZDCF.

\begin{table}
\begin{tabular}{lllll}
\hline
\hline
\multicolumn{1}{l}{Centroid Frequency}   & \multicolumn{2}{c}{$\rm \tau_1$ {[}d{]}}           & \multicolumn{2}{c}{$\tau_2$ {[}d{]}}               \\ 
\multicolumn{1}{l}{$f_L$ ($10^{-7}$ Hz)} & \multicolumn{1}{c}{Mean ($\mu$) } & \multicolumn{1}{c}{S.D ($\sigma$)} & \multicolumn{1}{c}{Mean ($\mu$) } & \multicolumn{1}{c}{S.D ($\sigma$)} \\ \hline
               $ 2.0$ (LF QPO)           &      57.84            &        3.37       &        116.30         &      7.83 \\ 
              $ 8.0$ (MF QPO)             &      14.37             &     0.48            &      28.93            &    0.61                \\ 
             $32.0$ (HF QPO)             &      3.99              &    0.045           &    7.00            &     0.00              \\ \hline
\hline                                         
\end{tabular}
  \caption{The mean and dispersion of the time values of the fundamental peak ($\rm \tau_1$) and the second peak ($\rm \tau_2$) for pure Lorentzians for 1000 MCS having $Q=30$ at the three test frequencies for the ACF.}
   \label{tab:ACFLor}
\end{table}

\subsection{The behavior of the ACF for pure Lorentzian processes}

\begin{figure}
\includegraphics[scale =0.6]{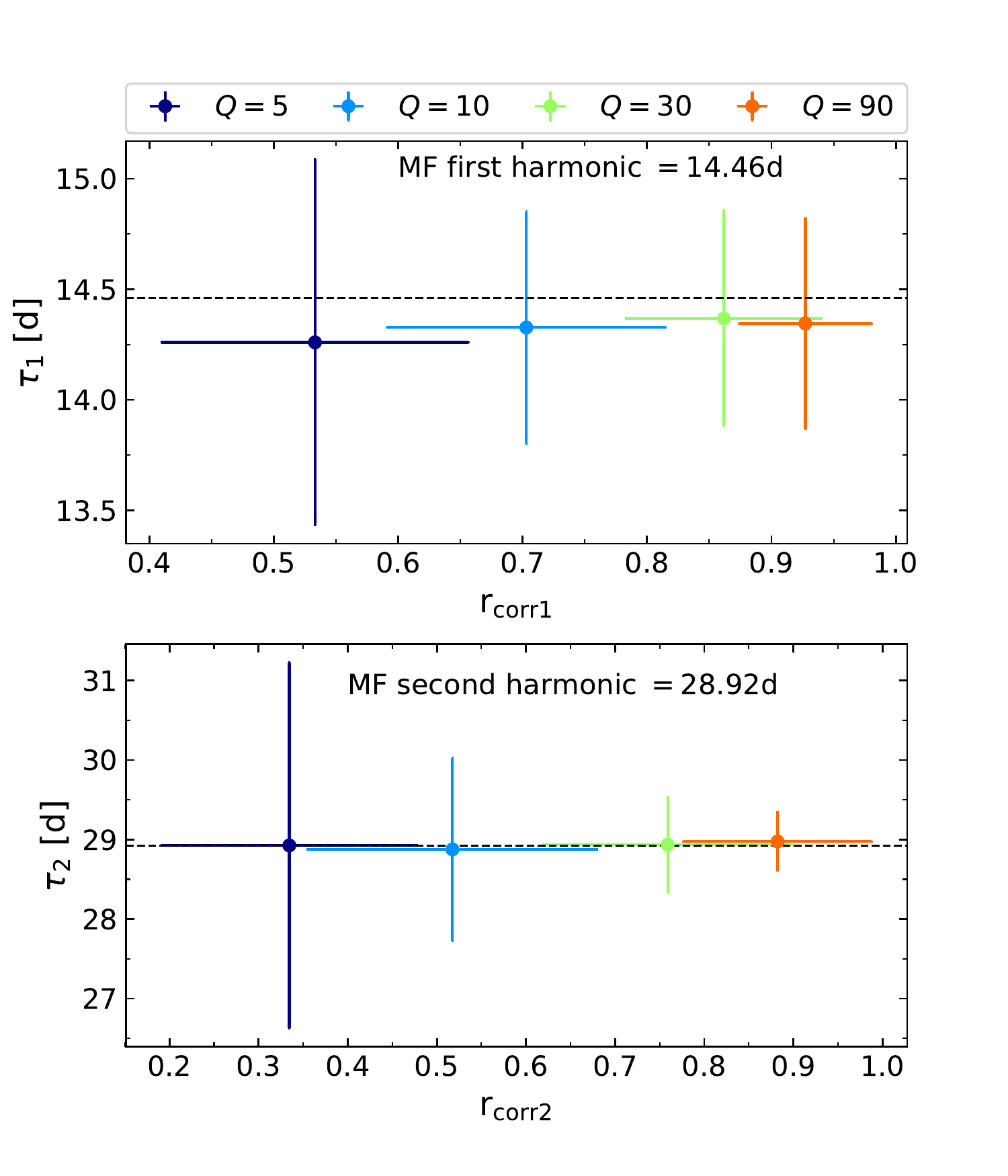}
\caption{The mean and dispersion of the times of the fundamental peak ($\rm \tau_1$) \& the second peak ($\rm \tau_2$) plotted against their corresponding correlation coefficient values for the MF Lorentzian signal obtained for 1000 MCS for few selected Lorentzian widths for the ACF.}
\label{fig:LorACF}
\end{figure}

In the ``ideal'' case of strictly periodic sinusoidal oscillations without the addition of any red noise variability, the ACF is a cosine, with correlation coefficient $\rm r_{corr} = 1$, at time lags corresponding exactly to the period and successive harmonics of the strictly periodic signal. In Lorentzian profiles, when the quality factor $Q$ is reasonably high ($Q \gtrsim 3$), there are still quasi-regularly spaced peaks occurring near the expected time lags for subsequent corresponding harmonics. To quantify the dispersion in peak time lag $\tau$ and $\rm r_{corr}$, we simulate 1000 light curves (which is sufficient to probe the 99.7\% confidence limit) for the pure Lorentzians spanning a range of quality factors from 1 to 120. We define the first peak as the maximum value of the ACF after the first negative-to-positive crossing and before the second positive-to-negative crossing. We determine the 99.9 per cent confidence range of $ \tau_1$, $ \tau_2$, $\rm r_{corr1}$ and $\rm r_{corr2}$ where the subscripts 1 and 2 correspond to the fundamental peak and the second peak (second harmonic) respectively. For values of $Q$ larger than $\sim20$, the dispersion in $\tau_1$ remains approximately constant as shown in Fig.~\ref{fig:LorACF}. Hence, we use a high value of $Q=30$ for all the following tests for our study. The dispersions of $\tau_1$ \& $\tau_2$ for the LF, MF \& HF QPOs are given in Table~\ref{tab:ACFLor}. We see that the 99.9 per cent limits in $\tau_1$ \& $\tau_2$ are within $\pm20$ per cent, $\pm10$ per cent and $\pm4$ per cent of the input value of $f_{\rm L}$ for the LF, MF \& HF cases respectively. We later use these 99.9 percent confidence contours as the limits within to search for the first and second harmonics when we search for QPOs mixed with broad-band red noise using the ACF at each different test frequencies. We note that the $99.9$ per cent limits of $\rm r_{corr1}$ range from 0.1 -- 1.0 in the LF case, while they are more constrained to values $>0.45$ in the MF \& HF cases. The limits of $\rm r_{corr2}$ range from 0.1--1.0 in the LF and MF cases, while they are constrained to values $> 0.4$ in the HF case.

\subsection{The behavior of the ACF for pure red noise power-law processes}

\begin{figure}
\includegraphics[scale =0.6]{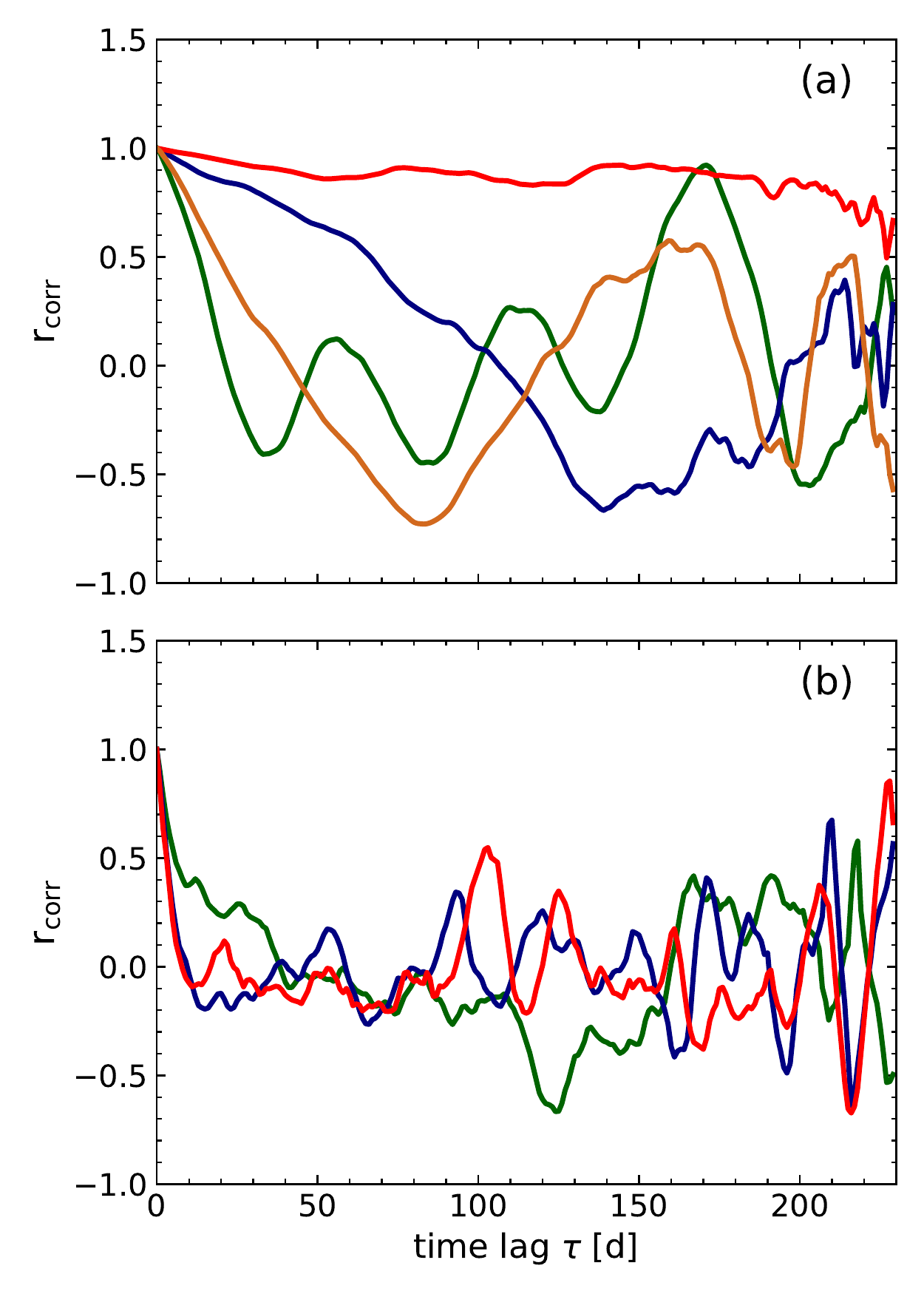}
\caption{The ACFs of pure red-noise light curves generated with (a) an underlying unbroken power-law PSD that has a power-law slope of $\beta = 2.2$, and with different random number seeds. Such a pure random stochastic red noise process causes broad bumps and wiggles in the ACFs. For example, consider the green line: the first peak occurs at time lag of $\sim$55 days, with successive peaks near 110 and 165 days. Such a signature is similar to that expected from a quasi-periodic signal and thus can possibly misintepreted as a period, (b) for an input PSD with a broken power-law model, with slope $\beta=2$ above a temporal frequency of {\bf $5.62 \times 10^{-7}$ } Hz breaking to $\gamma=1$ below it. Again, broad bumps and wiggles are common.}
\label{fig:RNACF}
\end{figure}

\subsubsection{Unbroken power law model}

We generate pure red noise light curves using an unbroken power-law PSD model for a wide range of slopes $\beta$ (0.4--3.0), in steps of ${\Delta}\beta$=0.2 for 1000 realizations at each step.

In Fig.~\ref{fig:RNACF}(a), we display the ACFs of a few selected light curves generated with the same power-law model, $\beta=2.2$, but with different random number seeds. We note the appearance of spurious broad bumps and wiggles in some ACFs; some of the peaks after the zero lag can attain values of $\rm r_{corr}$ as high as 0.8 or 0.98. For one particular realization, the green curve in Fig.~\ref{fig:RNACF}(a), the ACF has three broad bumps wherein the second and third peaks correspond to time lags roughly 2 and 3 times the delay of the first peak. This signature is similar to that expected for a quasi-periodic signal, but it was generated by a purely stochastic process.

As a preliminary exploration of false positives and the timescales over which they can occur, we adopt the following \textit{criteria for false positives: only the ACF that crosses zero level at $\rm r_{corr}=0.0$ and rises up to a non-zero value will register a positive detection at the time lag ($\tau_1$) with maximum value of correlation coefficient ($\rm r_{corr1}$). In this way, we have the false positives determined as a function of time lag.} Again, to register a second peak, the $\rm r_{corr}$ should drop from the maximum value corresponding to $\rm r_{corr1}$, cross zero and rise up again to peak at a non-zero value. That will register a second positive peak at the corresponding time lag $\tau_2$ at the second maximum correlation coefficient $\rm r_{corr2}$ after the zero lag. For illustration, in Fig.~\ref{fig:RNACF}(a), the green lines will register false positive detection at time lags corresponding to three peaks, while the orange and blue line will show false positives at time lags corresponding to the first and second peak after crossing zero. The ACF in red will not register any false positives at any time lag since it never crosses zero. Because the ACF and PDM points are very highly self-correlated in the presence of red noise, determining the effective number of independednt timescales is not straightforward (in contrast to, say, an evenly-sampled periodogram). Consequently, we emphasize that our false-alarm rate tests are conducted with the goal of detecting whether or not a pure red noise process can generate *at least one* false signal *at any frequency*. This means that our derived false-alarm probabilities are conservative in the sense that the likelihood of one single signal at one given frequency being real is even lower.

\begin{figure*} % "[t!]" placement specifier just for this example

\includegraphics[scale=0.65]{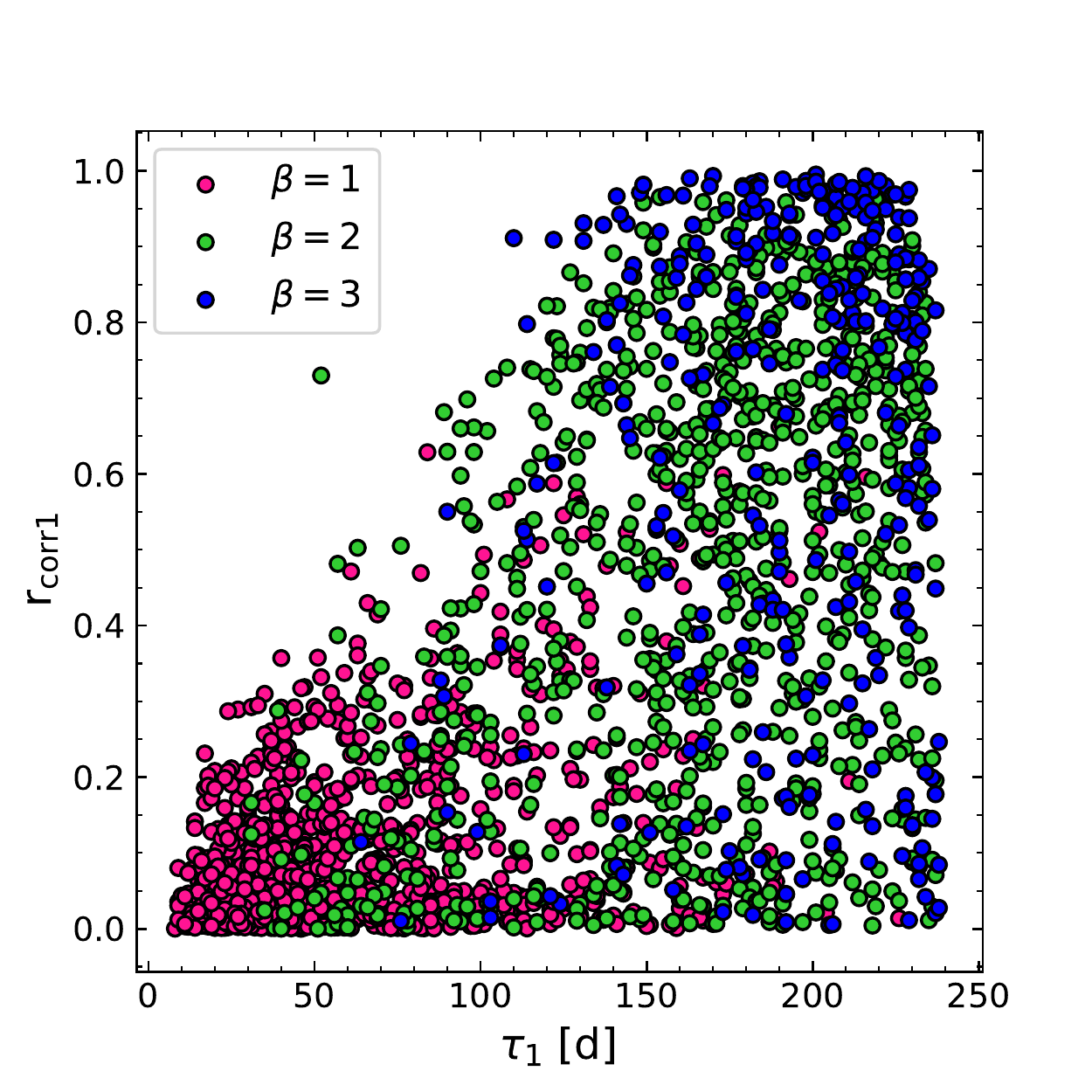}
\hspace*{0.5cm}
\includegraphics[scale=0.65]{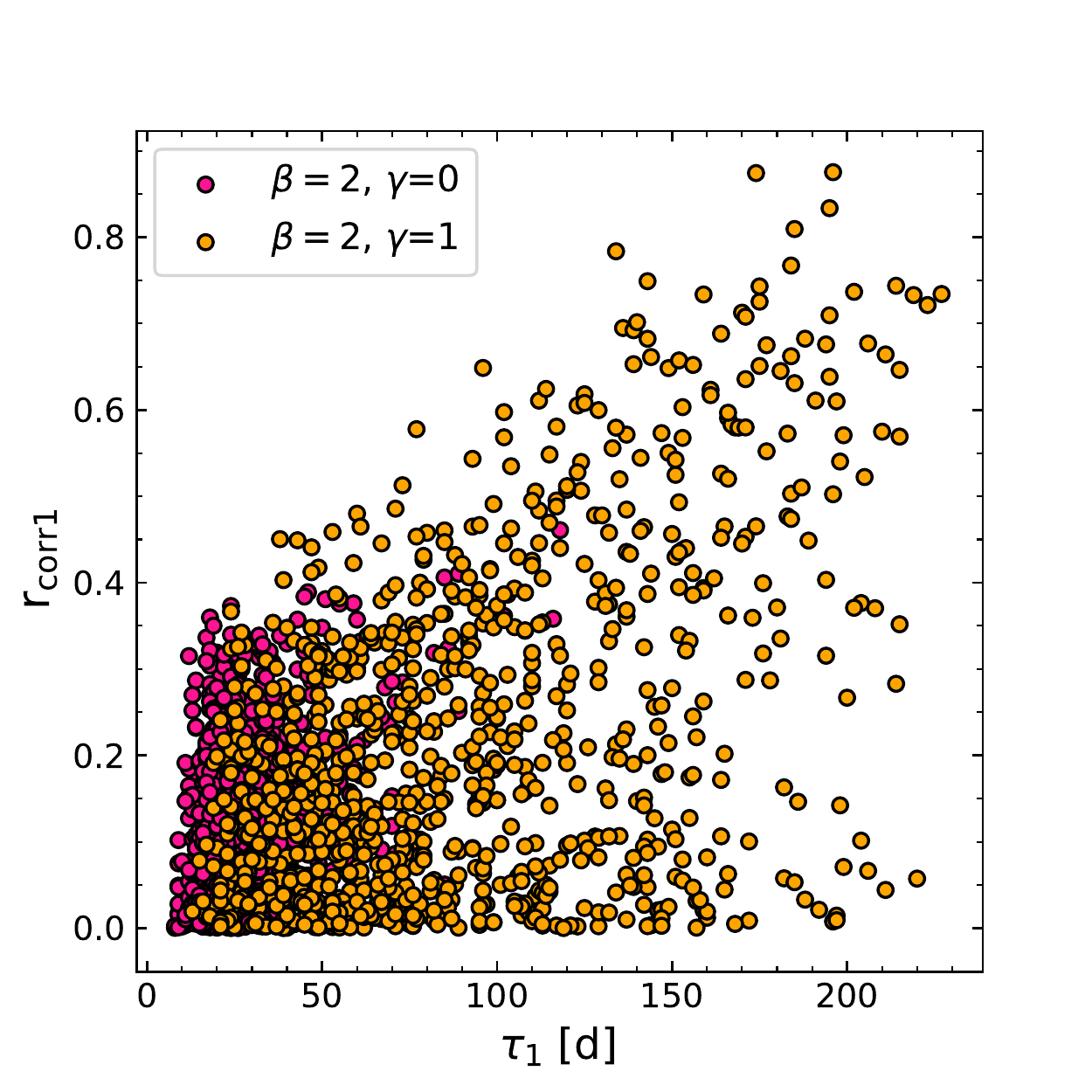}
\caption{The distribution of the time scales of the first peak after the zero lag against the corresponding correlation coefficient of false positives detected in ACF for 1000 simulations of pure red noise signals of (left) unbroken PL PSD model for few selected spectral index slopes (right) broken PL PSD model for $\beta = 2.0$ breaking to few selected low frequency slope ($\gamma$) below a temporal break frequency of $5.6 \times 10^{-7}$ Hz. }
\label{fig:acffapdistr}
\end{figure*}

In Fig.~\ref{fig:acffapdistr}, we plot the distribution of false-positive peaks as a function of timescale $\tau$ for four selected values of $\beta$. We note that the 99.9 per cent confidence distributions of $\rm r_{corr1}$ \& $\rm r_{corr2}$ of the first and second peak become wider with increasing slope: they have $\rm r_{corr1,2} <0.45$ for $\beta \lesssim 0.8$,0.6 and as the slope increases, it can range anywhere from 0 to as high as 0.9 or greater for $\beta \gtrsim 1.6$. This figure can be used to provide an estimate of the region of ($\tau_1$, $\rm r_{corr1}$) space where it is statistically likely or unlikely to encounter a false-positive peak. For a candidate peak at $\tau_1$, say, roughly 83 d $=  1/3$ of the duration, if $\beta=1.0$ (2.0), then only if $\rm r_{corr1}$ is greater than roughly 0.7 (0.8) will there be a strong likelihood ($>99.7$ per cent) that the signal is genuine. By $\tau_1=125$d, 1/2 of the duration, minimum values of $\rm r_{corr1}$ needed to confidently claim a genuine signal rise to $\sim$0.85 for $\beta = 2.0$.

Note that in Fig.~\ref{fig:acffapdistr}, for slopes 2.0 and steeper (green and blue points), the majority of the false positive peaks ($\tau_1$) occur at timescales longer than very roughly $1/3$ duration of the light curve: For $\beta =$ 1.0, 2.0, and 3.0, the fractions of false peaks residing at timescales of $> 1/3$ duration are $261/1000 =$  26.1 per cent, $790/857  =$  92.2 per cent, and $316/319  = $ 99.1 per cent, respectively. That is, simply excluding all peaks occurring at timescales longer than very roughly $1/3$ duration is a simple and effective way to reduce the total number of false-positive detections for relatively steep red-noise slopes. When considering timescales shorter than $\sim 1/3$ of the duration, false positive signals can also be minimized by excluding signal whose
values of $\rm r_{corr1}$ and $\rm r_{corr2}$ are less than $\sim$0.55 and 0.45, respectively.

We also try to look for peaks without the zero crossings, i.e., the ACF which crosses any non-zero level say at $\rm r_{corr}=0.2/0.3/0.4$ and rises up to a maximum value (i.e., $\rm r_{corr1} > 0.2/0.3/0.4$ respectively) is registered as a positive detection at the corresponding time lag ($\tau_1$). We find that even on choosing a non-zero crossing level, the red noise light curves produces significantly high percentage of false positives. For example, it shows false positives greater than 99.7 per cent (99 per cent) at $\beta \gtrsim 1.2$ (1.0) for the crossing level of $\rm r_{corr}=0.2$ (0.3).

\subsubsection{Broken power-law model}

\begin{figure}

\includegraphics[width=\linewidth]{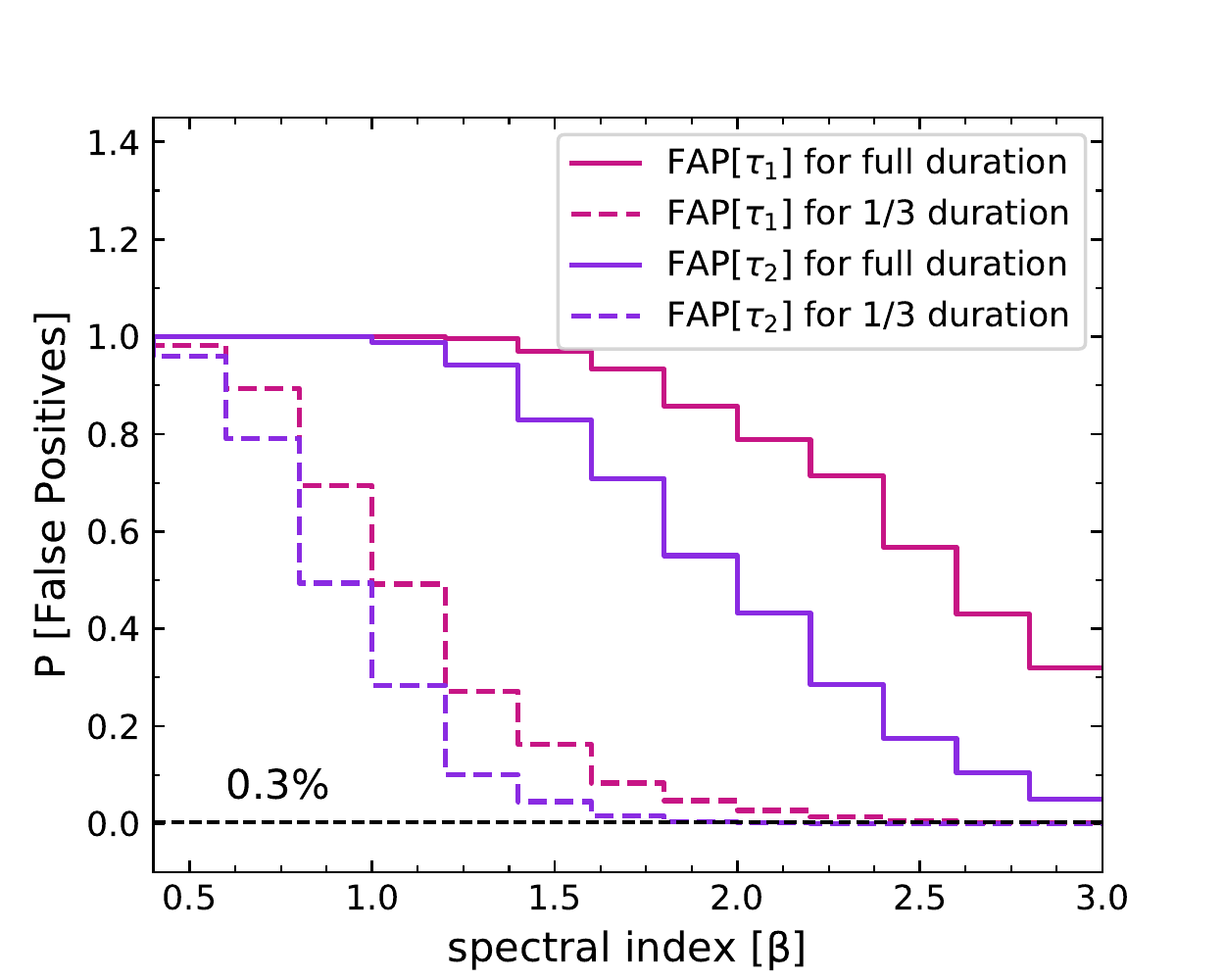}
\caption{The rate of false positives detected in pure red noise light curves having unbroken power law PSD model at different spectral index slopes for the first \& second peaks in the ACF while considering the full duration and then while considering just one-third of the total duration. }
\label{fig:acffap}
\end{figure}

\begin{figure*}
\includegraphics[width=\linewidth]{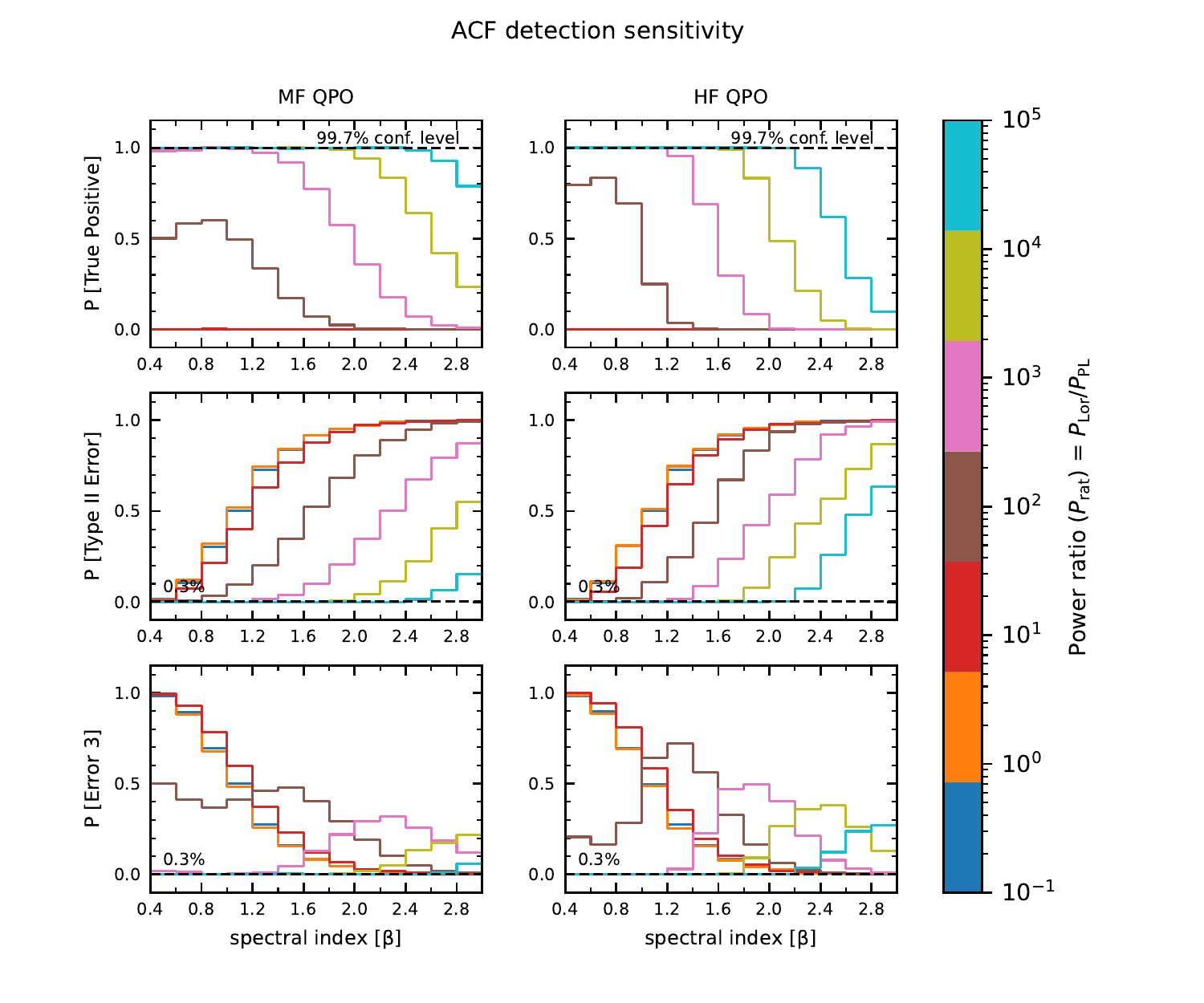}

\caption{Top panel: The probability of detecting the first harmonics of the periodic
          signal in the expected frequency range, Middle panel: The probability of detecting the first harmonics of the quasi-periodic
          signal in the wrong frequency range, Bottom panel: The probability of detecting false negative of the first harmonics of the quasi-periodic
          signal all as a function of power ratio log($P_{\rm rat}$) and
          red noise PSD power-law slope $\beta$ for the MF \& HF QPO mixed with broad-band red noise signals on using the ACF.}
\label{fig:QRNACF}
\end{figure*}

\begin{figure}

\includegraphics[width=\linewidth]{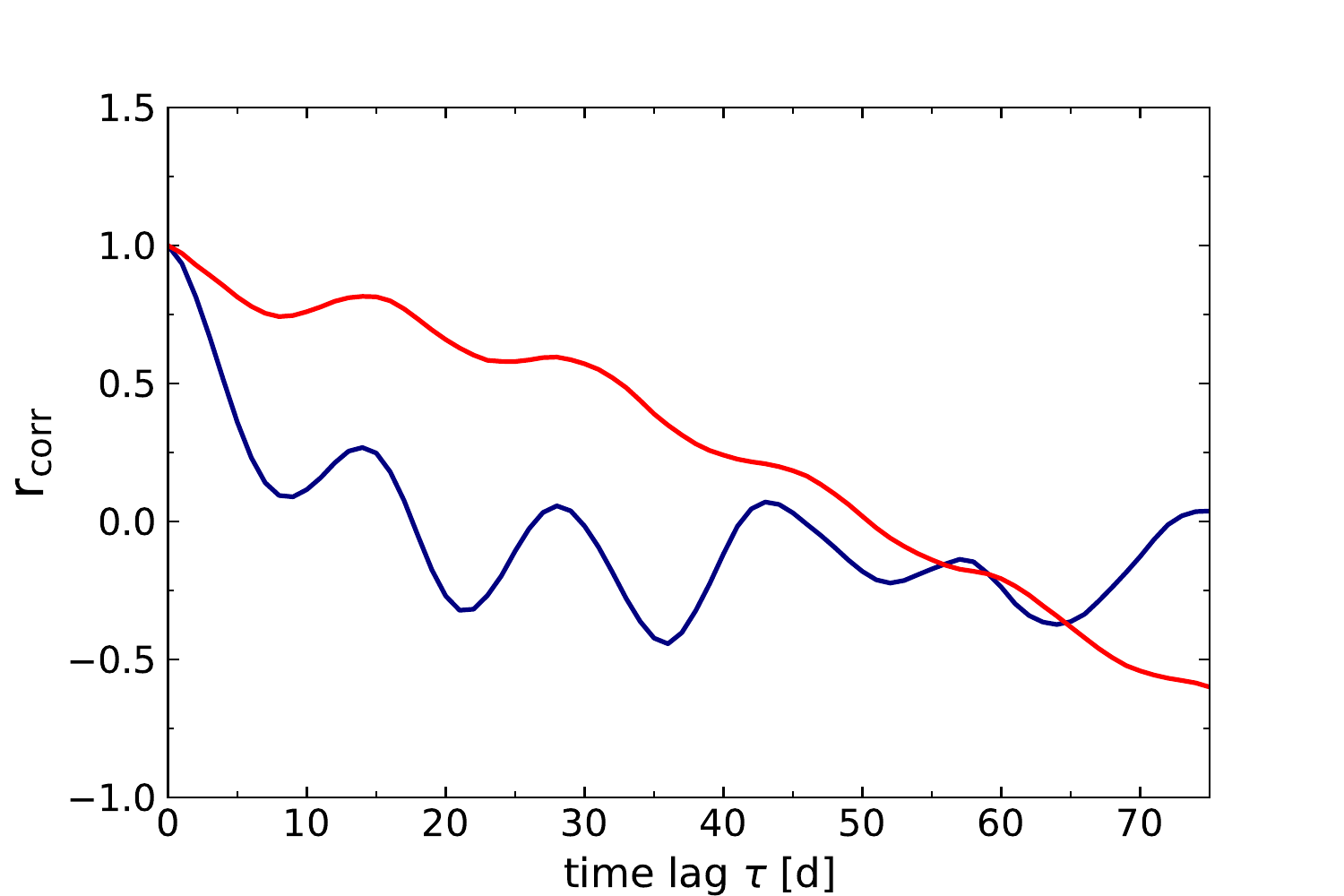}
\caption{Two selected ACFs for light curves corresponding to power spectra of MF QPO signal of log($P_{\rm rat})=2$ against the red noise continuum having $\beta = 1.8$. Mild features of quasi-periodicity are evident in the ACFs, but these ACF peaks do not satisfy our true positive detection criteria and would thus lead to false negative (red) and detection inaccuracy error (blue). }
\label{fig:acferr}
\end{figure}

We generate pure red noise light curves considering broken power-law models for 1000 realizations. We study it for just few selected values of high frequency spectral index slopes $\beta$ (1.0,2.0,3.0) breaking below a temporal mid-frequency of $5.6 \times 10^{-7}$ Hz to lower spectral index slopes of $\gamma$ (0.0,1.0,2.0).

In Fig.~\ref{fig:RNACF}(b), we plot the ACFs of a few selected light curves shown for a high frequency slope of $\beta=2.0$, breaking to low frequency slope of $\gamma=1.0$. Similar to the unbroken PL model, we see various bumps and wiggles in the ACF of broken PL model as well. The 99.9 per cent confidence distributions of $\rm r_{corr1}$ become wider with increasing slopes of $\beta$ and $\gamma$ independently: for $\beta=1.0$, 2.0 breaking to $\gamma=0.0$, has the upper confidence limit of $\rm r_{corr1}$ to be 0.35 and 0.45 respectively. When $\beta=3.0$, breaking to $\gamma=$ 1.0, 2.0, it has the $\rm r_{corr}$ range from 0 to 0.98.

In Fig.~\ref{fig:acffapdistr}, we plot the distribution of false positives of the first peak after zero lag for the case of $\beta=2.0$ breaking to $\gamma=0.0$ and $\gamma=1.0$. We see that for the case of $\beta=2.0$ breaking to a flatter spectrum, the majority fraction (99.2 per cent) of the candidate peaks ($\tau_1$) at timescales $<$ 83 d ($1/3$ duration) and with $\rm r_{corr1}<$ 0.4, while for breaking to a more steeper spectral slope of $\gamma = 1.0$, while it still has the many of the candidate peaks (55.9 per cent) below $1/3$ duration, it can still have false postives at longer timescales with high correlation coefficient values of $\rm r_{corr1} \gtrsim$ 0.8. Hence, we conclude that excluding peaks at timescales longer than very roughly $1/3$ duration of the light curve serves well in reducing the false-positive detections for broken-power law PSD models as well.

\subsubsection{False Alarm Probabilities}

In all discussions on errors, we consider the null hypothesis, H0, to be pure red noise variability, while the alternate hypothesis, H1, is a mixture of a QPO and red noise.

\textit{Type I Error (false positive): Expanding upon the discussion from above, we calculate how often H1 is incorrectly inferred true when H0 is intrinsically true} in the ACF of 1000 MCS of pure red noise light curves by using the selection criteria for false positive detections as described in $\S$3.2.1. We determine the probability of false positives of first and second peaks in the ACF across the time lag independently.

In Fig.~\ref{fig:acffap}, for the unbroken-PL model: When considering the full light curve duration, we notice that false positives of the first and second peak are detected with $\gtrsim 0.3$ per cent probability at all the tested values of pure red noise slope with the unbroken-PL model. For $ \beta \lesssim 1.4$, for example, an extremely high fraction --- roughly $99.6$ per cent --- of all simulated light curves yield atleast one false positive signal. We also note that at relatively steeper slopes of $\beta \gtrsim 2.0$, the fractions of the ACF crossing zero and producing a peak becomes much lower: the ACFs of the steepest slope red noise light curves are more wide, and most of the time they never cross zero to register a false positive based on our selection criteria. Hence, the probability of finding a second peak after the zero lag also becomes smaller towards the steeper spectral index slopes. 

As noted above in the preliminary findings, when we restrict ourselves to timescales less than one-third of the duration, the rate of false positives of the first and second peak after the zero lag in the ACFs of unbroken-PL model is $\gtrsim 0.3$ per cent only at $\beta \gtrsim$ 2.6, 1.8 respectively. But now, we see that less than 0.3\% of the simulations at all the spectral index slopes have correlation values corresponding to $\tau_1$ and $\tau_2$ greater than 0.45--0.5 . 

In the case of broken-PL PSD model: We tested for (i) $\beta=3.0$ breaking to three values of $\gamma=0.0$, 1.0, 2.0; (ii) $\beta=2.0$ breaking to two values of $\gamma=0.0$, 1.0 and (iii) $\beta=1.0$ breaking to $\gamma=0.0$, below a temporal mid-frequency of $5.6 \times 10^{-7}$ Hz. When considering the full duration of the light curve in the ACF, we see that the probability of false positive of the first and second peak is $> 0.3$ per cent for all the three cases. In fact greater than $99.7$ per cent of the time, it is probable to get atleast a single peak at some time lag for the cases of $\beta=1.0$, 2.0, 3.0 breaking to $\gamma=0.0$, 1.0. When we restrict to timescales $<1/3$ duration in the ACF, we still have the false positives of the first and second peak $> 0.3$ per cent, but the corresponding correlation values are now restricted to $\rm r_{corr1,2} < 0.55$ for all the three cases. We note that particularly in the case of $\beta =1.0$ breaking to $\gamma=0.0$, it reaches $>99.7$ per cent probability of false detection for the 1000 simulations even after considering just one-third of the duration, but the maximum value of correlation coefficient that can be reached is $\rm r_{corr1}<0.35$. 

% Results on using ACF of QPOs mixed with red noise
Therefore, from the above tests we conclude that by restricting to searching for peaks in the ACF at lags less than one-third of the lightcurve duration and considering the first peak after the zero lag to have the correlation coefficient greater than 0.45--0.5, one can avoid the detection of false-positives and have the false positive rates of the period of the signal to be significantly low ($<0.3\%$) across all the spectral index slopes. 

\subsection{ACFs for QPOs mixed with red noise}

We perform MCS for the sum of a $Q=30$ Lorentzian and an unbroken power-law red noise continuum, with slopes spanning 0.4--3.0 in steps of ${\Delta}\beta=0.2$, and power ratios log($P_{\rm rat}$) ranging from $-$1 to +5 in steps of ${\Delta}$log($P_{\rm rat}) = 1$  for $N=1000$ light curves at each step.  

\textit{True-positive detections:} The criteria for true positive detections are as follows: we register true positive detections for the first or second harmonics of the QPO independently only when they fall in their respective expected ranges of time lags as derived from the ``ideal'' case of a $Q=30$ Lorentzian (no red noise) and consider just one-third of the duration in the ACF while searching for the peaks. The corresponding correlation coefficient of the first peak also needs to have $\rm r_{\rm corr1} > 0.45$, since we need to check for the $99.7$ per cent significance level to claim a true detection and to simply rule out the possibility of the peaks being due to the red noise continuum having a broken power-law model form.

We note that we do not perform the test for the LF QPO signal mixed with red noise since the distribution of $\rm r_{\rm corr1}$ even for true lorentzian ranges from 0.1--1. It is too low in order to statistically eliminate the possibility of true positive detection from false positives (e.g, red noise with broken power law model), since we cannot restrict it along the correlation coefficient values.

The detection rate of true positives for both the first and second harmonics decreases as log($P_{\rm rat}$) decreases and/or as the power-law slope $\beta$ increases; this behavior holds across MF, and HF QPOs, as illustrated in Fig.~\ref{fig:QRNACF}. In order to ensure a 99.7 per cent or higher reliability for detection of a QPO, one typically requires a very high power ratio. To register true positives of $\tau_1$ with 99.7 per cent significance having $\rm  r_{ corr1}>0.45$: in the MF case, log($P_{\rm rat}$) must be 5, or 4 for values of $\beta$ $\lesssim 2.4$, or $\sim 1.8$ respectively. In the HF case, log($P_{\rm rat}$) must be 5, 4, or 3 for values of $\beta$ $\lesssim 2.2$, $\sim1.6$, or $\sim1.2$ respectively. To register true positives of $\tau_2$ having $\rm r_{ corr2}>0$: for MF case, log($P_{\rm rat}$) must be 5, or 4 for values of $\beta$ $\lesssim 2.4$, or $\sim 1.6$ respectively (where $\rm  r_{corr2}$ is always greater than 0.2). In the HF case, log($P_{\rm rat}$) must be 5, 4, or 3 for values of $\beta$ $\lesssim 2.0$, $\sim1.6$, or $\sim1.0$ respectively (where $\rm  r_{corr2}$ is always greater than 0.4).

\subsubsection{False negatives and detection inaccuracy errors}

\subsubsection*{Type II Error (false negative):H1 is inferred false incorrectly when a QPO is present.} We determine the number of times the first and second harmonics are never triggered positive (at \textit{any} timescale) in the ACF of the light curves determined independently across the time lag on using the true-positive detection criteria.

In a light curve containing a QPO mixed with red noise, to account for false negatives of the first peak, we determine the number of times not even a single peak is detected after the zero lag in the ACF (i.e, it never crosses $\rm r_{corr} = 0$ and then rises above it) of 1000 simulations; to determine the false negatives of the second peak, we determine the number of times the ACF never crosses $\rm r_{corr} = 0$ and rises above it after the first peak at $\tau_1$ of 1000 simulations determined independently of the time-scale of the first lag. For illustration, in Fig.~\ref{fig:acferr}, we plot two selected ACFs for light curves corresponding to power spectra with MF QPO signal of log($P_{\rm rat})=2$ against the red noise continuum having $\beta = 1.8$. Mild features of quasi-periodicity are evident in the ACFs. The ACF shown in red in Fig.~\ref{fig:acferr} has no peaks after crossing $\rm r_{corr}=0$. It does not satisfy our positive detection criteria and would thus lead to a false negative.

Particularly for the steepest power-law slopes --- $\beta > 2$ --- the probability of obtaining a false negative is quite high for all but the largest values of $P_{\rm rat}$. Even for such large values of log($P_{\rm rat}$) and large values of $\beta$, the dominating red noise causes the ACF central peak to simply not reach zero and completely overwhelms the cosine-like behavior from the QPO.

For the MF QPO case, negligible false negative probabilities ($<0.3$ per cent) for the first and second harmonic are obtained only for values of log($P_{\rm rat}$) $\sim$ 2, 3, 4, and 5 at $\beta$ $\lesssim$ 0.6, 1.2, 1.8, and 2.4, respectively.
By the time one reaches log($P_{\rm rat}$) $\sim$ 1, a significant percentage ($\sim99.7$ per cent) of simulations yield false negatives for the first (second) harmonic at $\beta$ $\gtrsim$ 2.8 (2.4).
For the HF case, it is a similar story: negligible false negative probabilities for the first and second harmonic are obtained  only for values of log($P_{\rm rat}$) $\sim$ 2, 3, 4, and 5 at $\beta$ $\lesssim$ 0.8, 1.2, 1.6, and 2.2, respectively.
False negative probabilities for the MF and HF cases are plotted in Fig.~\ref{fig:QRNACF}.

\subsubsection*{Detection inaccuracy error (Error 3): H1 is intrinsically true and also registered as true,
but the QPO do not satisfy the true positive detection criteria.} We define a detection inaccuracy error as follows: when a light curve containing a QPO mixed with red noise, produces a positive detection in the ACF, but (i) when it has the first (second) harmonic not within the ranges of time lags as derived from the ``ideal'' case of a $Q=30$ Lorentzian (no red noise) or (ii) when the correlation coefficient of the first harmonic within the expected time lag has $\rm r_{corr1}<0.45$. For example, in Fig.~\ref{fig:acferr}, the ACF in blue, which corresponds to light curve having a MF QPO signal of log($P_{\rm rat})=2$ against the red noise continuum having $\beta = 1.8$, has the first peak outside the frequency range expected for the MF QPO and hence will be registered as Error 3.

The probability of detecting a QPO at the wrong timescale increases with decreasing $P_{\rm rat}$. However, with increasing power-law slope $\beta$, the probability of successfully detecting a QPO at \textit{any} timescale (within the correct range or outside of it) decreases (especially for log($P_{\rm rat}) \lesssim 3$), hence the probability of detecting a QPO at the wrong timescale also decreases.
 
In general, we find that the probability of detecting the QPO (first peak) at the wrong timescale is very low ($\lesssim 0.3$ per cent) only for values of log($P_{\rm rat}$) $\gtrsim$ 4--5 and only when avoiding the steepest values of $\beta$ tested.
When log($P_{\rm rat}$) $\sim$5, detection inaccuracy probabilities are $<$ $\sim 0.3$ per cent for nearly all values of $\beta$ tested: the probabilities exceed $\sim 0.3$ per cent only for $ \beta$ $\sim 3.0$, and $\gtrsim2.4$ (MF, and HF, respectively).
However, when log($P_{\rm rat}$) $\sim$4, detection inaccuracy probabilities are $<$ $\sim 0.3$ per cent only when $\beta$ is $\lesssim$ 1.8, and 1.6 (MF, and HF, respectively).
When log($P_{\rm rat}$) $\sim$3, detection inaccuracy probabilities are $<$ $\sim 0.3$ per cent only when
$\beta$ is $\lesssim$ 1.2 for the HF case. 

The probabilities of obtaining a detection inaccuracy error for the MF and HF cases are plotted in Fig.~\ref{fig:QRNACF}.

\section{PDM ANALYSIS FOR PERIOD SEARCHING}
\begin{figure*}
\includegraphics[height= 14cm, width =18cm]{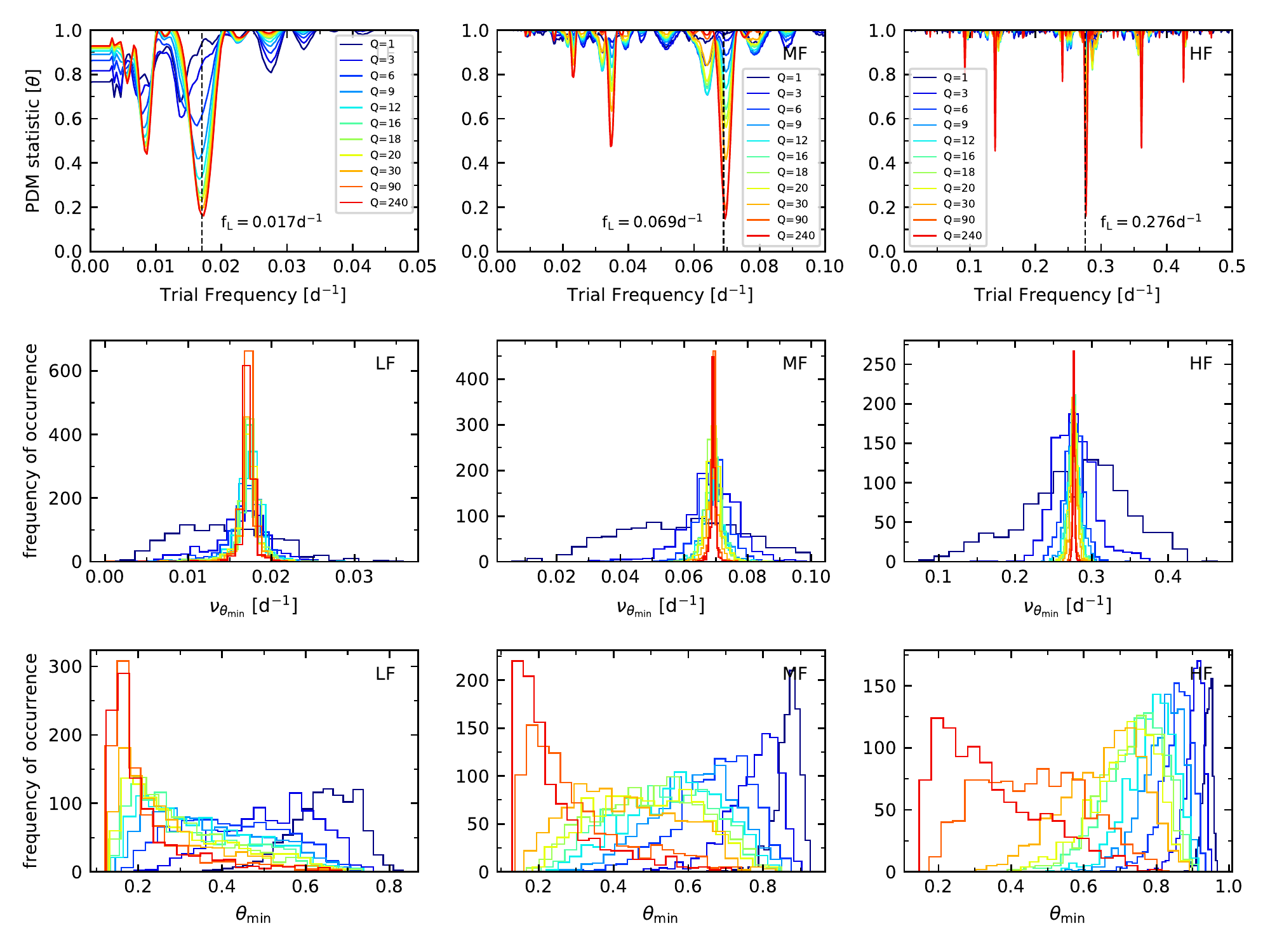}

\caption{Top panel: The PDM periodogram showing the characteristic dip at the expected frequency for one realization for different widths of Lorentzian profile at the three test frequencies, Middle panel: The histogram of the frequency $\rm \nu_{{\theta}min}$ [$\rm d^{-1}$] corresponding to the minimum statistic value $\rm \theta_{min}$, Bottom panel: The histogram of minimum statistic value $\rm \theta_{min}$ for different widths of Lorentzian profile at the three test frequencies for 1000 MCS when using the PDM.}
\label{fig:QPDM}
\end{figure*}

The Phase Dispersion Minimization (PDM) method is a widely-used numerical technique for searching for periodic pulsations in the fluxes of various astronomical objects \citep{1978ApJ...224..953S}. The method is suitable especially when the waveforms are non-sinusoidal. Consider a time series of
$N$ observational points represented as $(x_i, t_i)$, where $x_i$ is the value of the flux or count rate at time $t_i$. The variance of the flux is given by

\begin{equation}
\sigma^2 = \frac{\sum (x_i - \bar{x})^2}{N-1}, (i = 1,...,N)
\end{equation}

where $\bar{x}$ is the mean of the fluxes. The light curve is divided into a series of $M$ phase bins according to the trial period $\phi$ and each contains $n_j$ data points that are similar in phase. The variance of amplitude in each bin is computed, defined by

\begin{equation}
s^2_j = \frac{\sum (x_{kj} - \bar{x_j})^2}{n_j-1}, (j = 1,...,M)
\end{equation}

where $x_{kj}$ represents the $k$th data point in the $j$th phase bin. The overall variance for all the phase bins is computed by

\begin{equation}
s^2 = \frac{\sum (n_j-1)s^2_j}{\sum n_j - M}, (j = 1,...,M)
\end{equation}

One measures the ratio of the sample variance to the overall variance of the lightcurve, defined as the periodogram statistic $\theta$, given by $\theta = s^2/\sigma^2$. It gives the measure of scatter of the sample variance about the mean of the light curve. For a true period, the scatter of the sample variance about the mean will be small, and hence we expect $\theta$ to approach zero at that frequency. For frequencies that do not correspond to a true period, the scatter of the sample variance about the mean will be large where $s^2 \approx \sigma^2$ and the test statistic $\theta$ approaches 1. One can plot the test statistic $\theta$ at each trial frequency and determine the local minimum value $\theta_{\rm min}$, which indicates the frequency corresponding to the least scatter about the mean light curve. It was demonstrated that the test statistic $\theta$ actually follows a beta distribution (\citealt{1997ApJ...489..941S}) from which the significance of any detected pulsation can be obtained.
  
However, applications in astronomy have been primarily limited to situations where the underlying noise is Poisson (i.e., white noise). The performance of the PDM has not yet been empirically tested for situations where stochastic red noise backgrounds exist --- it is not clear that simply identifying the local minimum value $\theta_{\rm min}$ automatically indicates that a strictly/quasi-periodic signal exists at that frequency.

\begin{table}
\begin{tabular}{lccll}
\hline
\hline
      & \multicolumn{1}{l}{\begin{tabular}[c]{@{}l@{}}upper limit on $\rm \theta_{min}$\\                ($99.9\%$)\end{tabular}} & \multicolumn{1}{l}{\begin{tabular}[c]{@{}l@{}}Range of $\nu_{\rm \theta_{min}}$ [$\rm d^{-1}$]\\                       ($99.9\%$)\end{tabular}} &  &  \\ \hline
LF QPO & 0.684                                                                                                                      & 0.0096--0.0216                                                                                                                                           &  &  \\ \hline
MF QPO & 0.805                                                                                                                        & 0.0608--0.0754                                                                                                                                           &  &  \\ \hline
HF QPO & 0.883                                                                                                                      & 0.262--0.293                                                                                                                                           &  &  \\ \hline
\hline
\end{tabular}
  \caption{The 99.9 per cent contour limits of $\theta_{\rm min}$ and trial frequency $ \nu_{\theta_{\rm min}}$ for the Lorentzians at the three test frequencies.}
   \label{tab:PDMLor}
\end{table}
\subsection{PDM test for Lorentzian profile:}

We first consider the ``ideal'' case of QPOs, as represented by simple Lorentzians centered at each of our three test frequencies and spanning a range of values of quality factor $Q$.

For each values of $Q$, we simulated 1000 light curves, again using our ``baseline'' sampling pattern, measure their PDMs, and plot the test statistic $\theta$ determined at each trial period versus the frequency; sample PDMs for the LF, MF, and HF are plotted in Fig.~\ref{fig:QPDM}. Each show a characteristic dip ($\theta_{\rm min}$) at the expected frequency ($\nu_{\theta_{\rm min}}$), plus harmonics; relatively higher values of $Q$ yield narrower dips and lower values of $\theta_{\rm min}$. We plot the distributions of $\theta_{\rm min}$ and $\nu_{\theta_{\rm min}}$ for LF, MF and HF QPO in Fig.~\ref{fig:QPDM}.

The standard deviation of $\nu_{\theta_{\rm min}}$ decreases as $Q$ increases. Extremely coherent signals --- values of $Q$ of order 10--20 --- are typically needed to attain values of $\theta_{\rm min}$ below $\sim0.3$ for the LF and MF cases; for the HF case, values of $Q$ closer to 90 are required.

Henceforth, we consider Lorentzians with values of $Q=30$. To delineate the region of frequency--$\theta$ space where the ``ideal'' signal occurs, we determine the 99.9 per cent confidence regions from the MCS denoting the distributions of $\theta_{\rm min}$ (upper limits) and $\nu_{\theta_{\rm min}}$; those upper limits/ranges are listed in Table~\ref{tab:PDMLor}.

We will later use these confidence ranges as the limits within which to search for signatures of QPOs when we consider mixtures of a QPO and red noise while using the PDM.

\begin{figure}
\includegraphics[width=\linewidth]{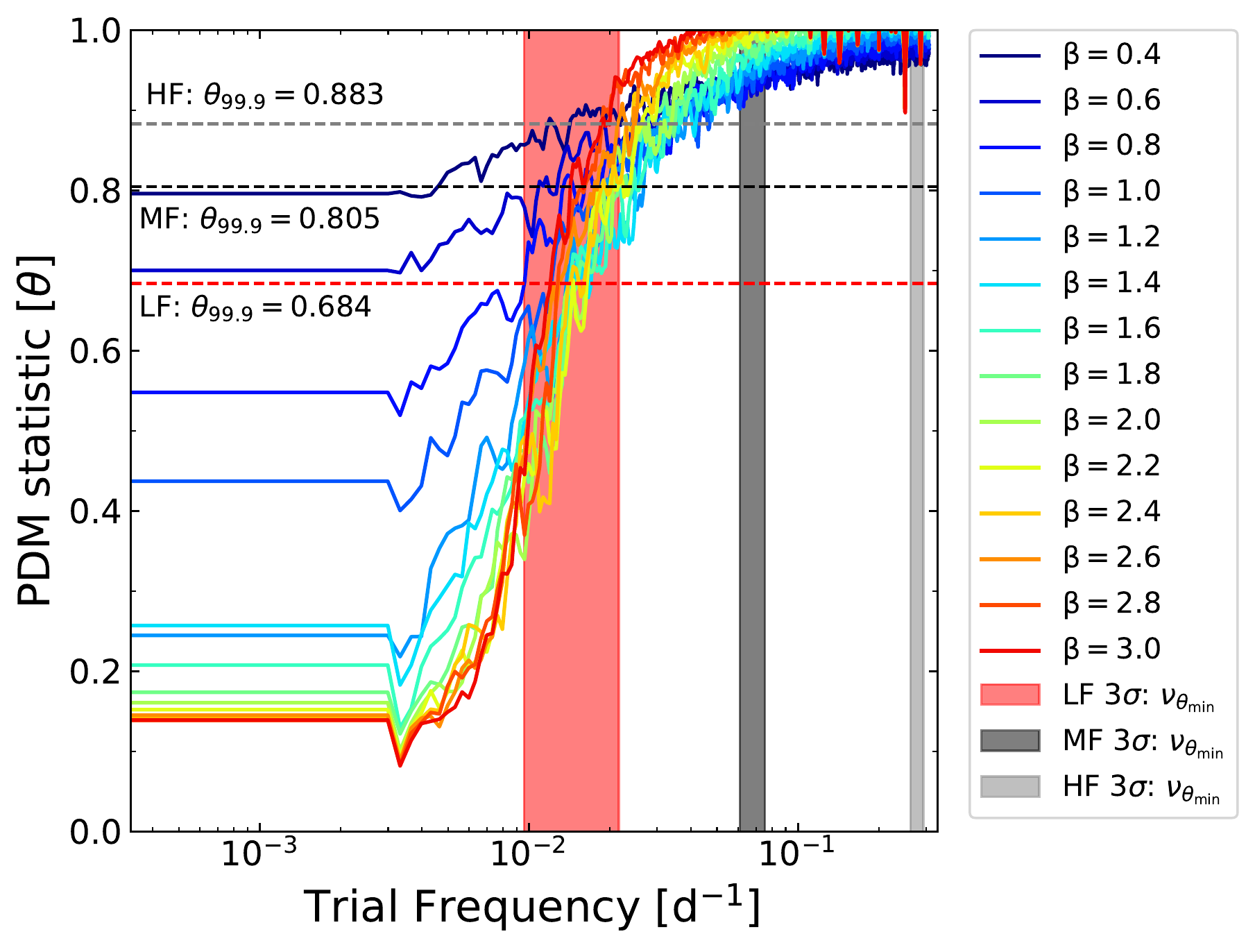}

\caption{The overplot of the 99.9 per cent lower limit of the PDM statistic value $\rm \theta$ at each test frequency that can be caused due to pure red noise light curves having evenly sampled data for broad-band spectral index slopes against the $99.9$ per cent contour limits of $\nu_{\theta_{\rm min}}$ and the upper limit on the $\theta_{\rm min}$ determined for pure Lorentzian signals at the three test frequencies for 1000 MCS within which we expect the characteristic dip to occur for QPOs mixed with red noise. }
\label{fig:RPDM}
\end{figure}

\subsection{The behavior of the PDM for pure red noise processes}

\subsubsection{Unbroken and broken power-law models}

We again generate pure red noise light curves using an unbroken power-law PSD model for a wide range of slopes $\beta$ (0.4--3.0), in steps of ${\Delta}\beta$=0.2, simulating 1000 light curves at each step, and measuring their PDMs.

We determine the periodogram statistic $\theta$ across all the possible test frequency for each light curve. The resulting 99.9 per cent confidence lower limits on $\theta$ at each trial frequency are plotted in Fig.~\ref{fig:RPDM} for the different spectral index slopes. For all power-law slopes, $\theta$ is typically $>$0.9 across a wide range of frequencies, and even $>$0.8 down to $\sim$0.02 d$^{-1}$ (i.e., the scatter variance is generally large above this frequency). At the lowest frequency bins however (at frequencies corresponding to timescales longer than $\sim$ 1/3 of the duration), $\theta$ plunges to 0.6 or lower, with relatively steeper PSD slopes reaching values of $\theta$ well below 0.2 and these local minimum due to pure red noise can be misintepreted as quasi-periodic signals. In fact, we note that the fraction of pure red noise-only trials that can attain a value of $\theta_{\rm min} < 0.6$ is greater than 99.7 per cent at $\beta \geq 2.6$ and it is less than 0.3 per cent at $\beta \leq 0.8$. However, after excluding the frequencies towards the lowest bin (at about $1/3$ of the duration), we see that the fraction of pure red noise-only trials that can obtain a $\theta_{\rm min} < 0.6$ is less than 0.3 per cent almost across all the tested beta values.

We repeat the test for pure red noise light curves considering broken power-law models for 1000 realisations. We study it for just few selected values of high frequency spectral index slopes $\beta$ (1.0, 2.0, 3.0) breaking below a temporal mid-frequency of $5.6 \times 10^{-7}$ Hz to lower spectral index slopes of $\gamma$ (0.0, 1.0, 2.0). Similar to the unbroken power-law models, we see that towards the lower frequency bins $\theta$ gets minimized to $< 0.6$. The fractions of red noise light curves that could reach $\theta_{\rm min} < 0.6$ is less than $1$ per cent when any value of $\beta$ breaks to flatter slope of $\gamma =0.0 $, but the fractions get higher as $\beta$ increases ($> 40$ per cent) on breaking to $\gamma =1.0$ or 2.0. After excluding the lowest bin frequencies (corresponding to $\sim$ $1/3$ duration), we see that the fraction of pure red noise-only trials that can obtain a value of $\theta_{\rm min} < 0.6$ is less than 1 per cent for all the combinations of broken power-law PSD model.  

\textit{It is very clear that low values of $\theta$ in a given PDM do not automatically mean that a QPO is present,} since red noise can also generate such low values of $\theta$.  In particular, however, one must take into account the frequency regime where the minimum occurs before attempting to decide if a given minimum corresponds to a genuine QPO. For example, in Fig.~\ref{fig:RPDM}, we see that at timescales $\sim$ $1/7$--$1/12$th of the total duration, the 99.9 per cent lower limit of $\theta_{\rm min}$ due to pure red noise ranges between 0.8--0.88, which are below the 99.9 per cent upper limit of $\theta_{\rm min}$ that can be reached by the HF pure Lorentzian signals but they do not occur within their expected frequency range. In fact, we note that the $99.9$ per cent lower limit of $\theta_{\rm min}$ caused due to pure red noise never intersects with the $99.9$ per cent search contour limits obtained for the MF \& HF pure Lorentzian signals because it is always greater than 0.89 around those frequency bins. On the other hand, the pure red noise light curves (e.g., $\beta<1.2$) can get $\theta \lesssim 0.6$ within the 99.9 per cent contour limits of $\nu_{\theta_{\rm min}}$ expected for a pure LF Lorentzian signals. Hence, we will not to be able to statistically distinguish a LF QPO signal mixed with red noise causing the minimized $\theta$ value. 

  \begin{figure}

\includegraphics[width=\linewidth]{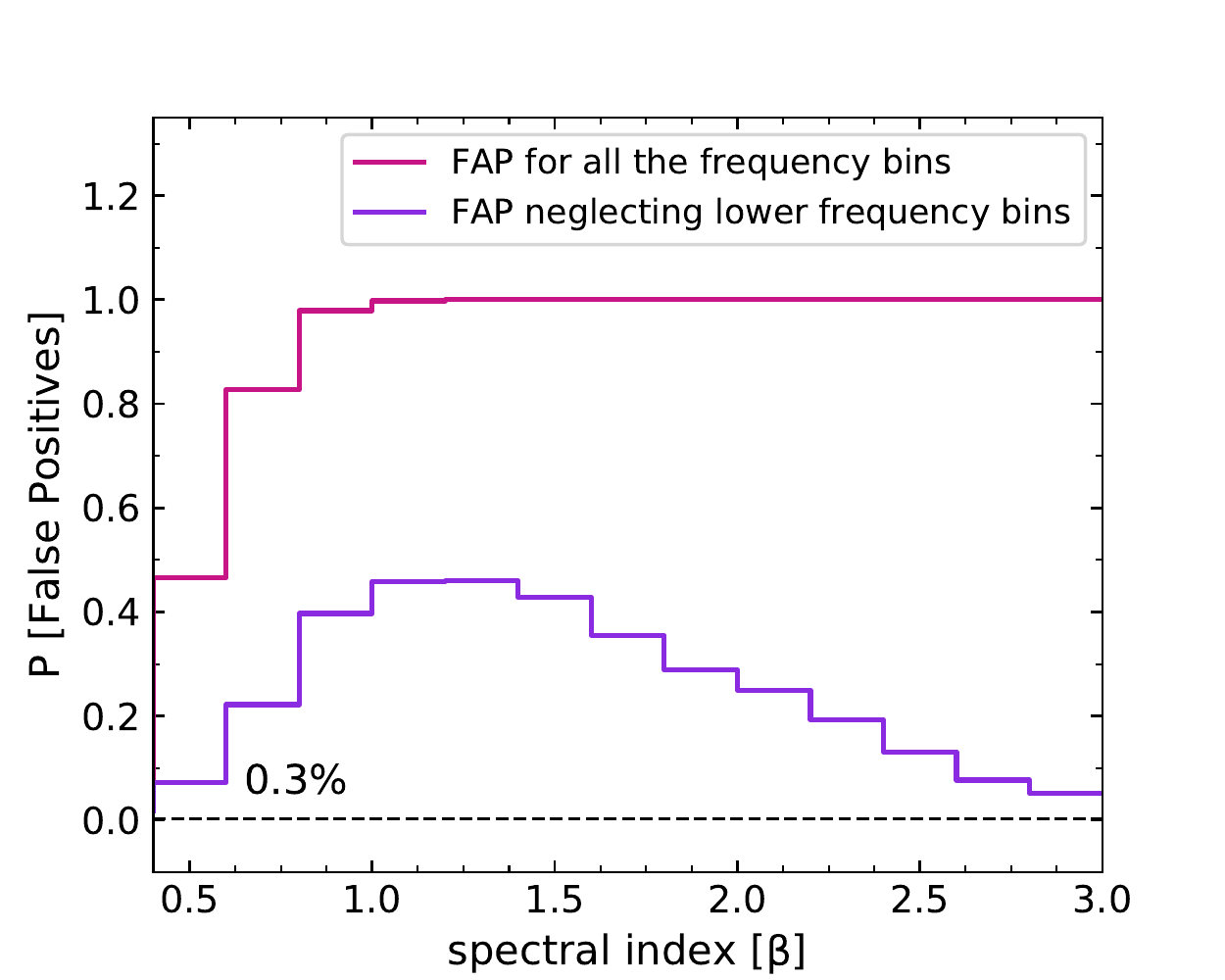}
\caption{The rate of false positives detected in pure red noise light curves having unbroken power law PSD model at different spectral index slopes while using the PDM tested for all the trial frequency bins and then neglecting the lower frequency bins corresponding to timescales $>1/3$ duration. }
\label{fig:pdmfap}
\end{figure}

\begin{figure}

\includegraphics[width=\linewidth]{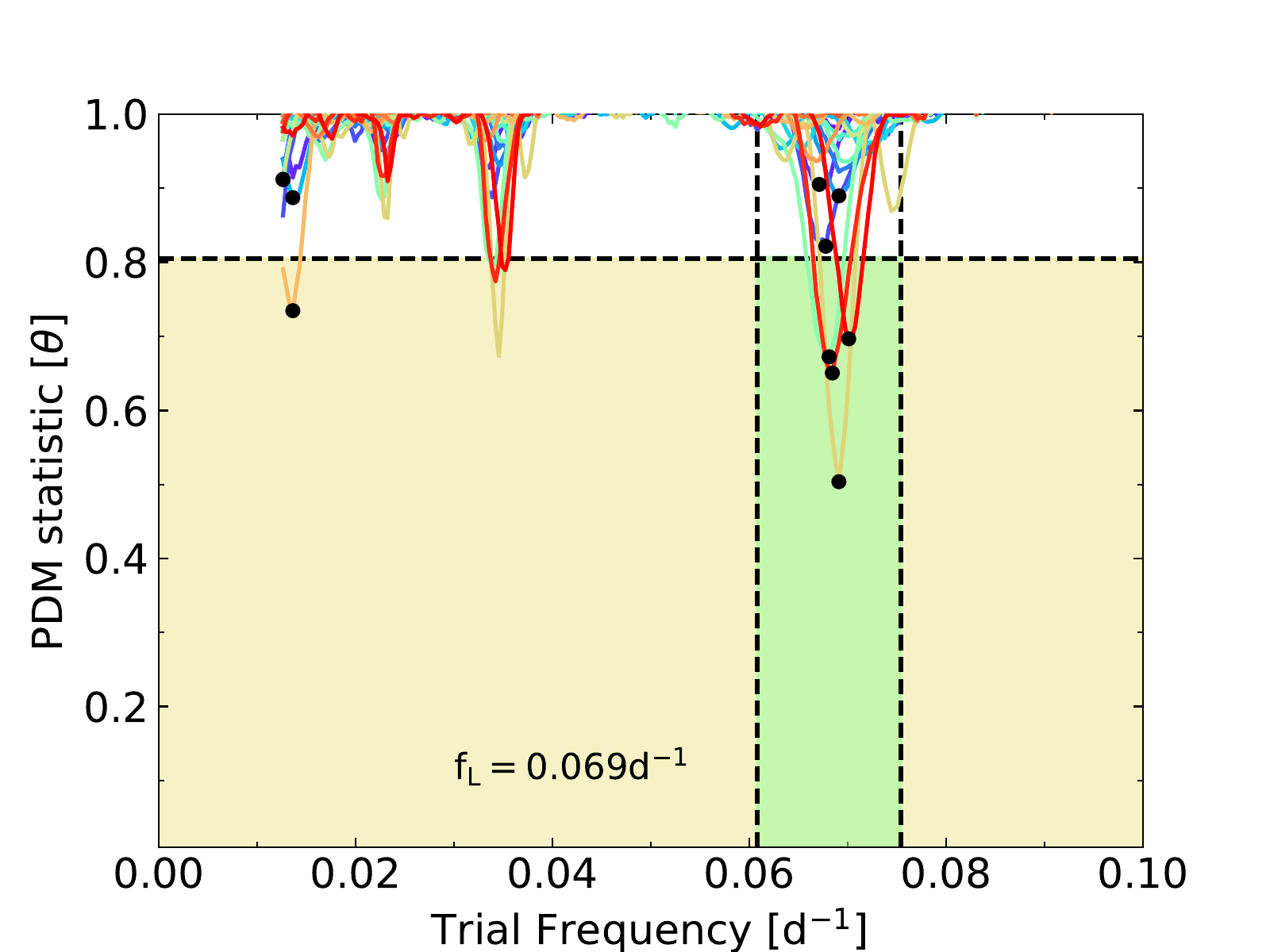}
\caption{The PDM statistic $\theta$ measured at each trial frequency bin for light curves having a MF QPO signal mixed with red noise having unbroken power-law model of slope $\beta = 2.6$ of log($P_{\rm rat}) = 2$, on using the PDM while neglecting the lower frequency bins corresponding to $\sim 1/3$ duration. The black points represent the lowest minimum ($\theta_{\rm min}$) for a given PDM periodogram and the green shaded region is the $99.9$ per cent contour limits of the ``ideal signal'' in the $\theta$--frequency space expected for the MF QPO signal. }
\label{fig:pdmerr}
\end{figure}

\begin{figure*}
\includegraphics[width=\linewidth]{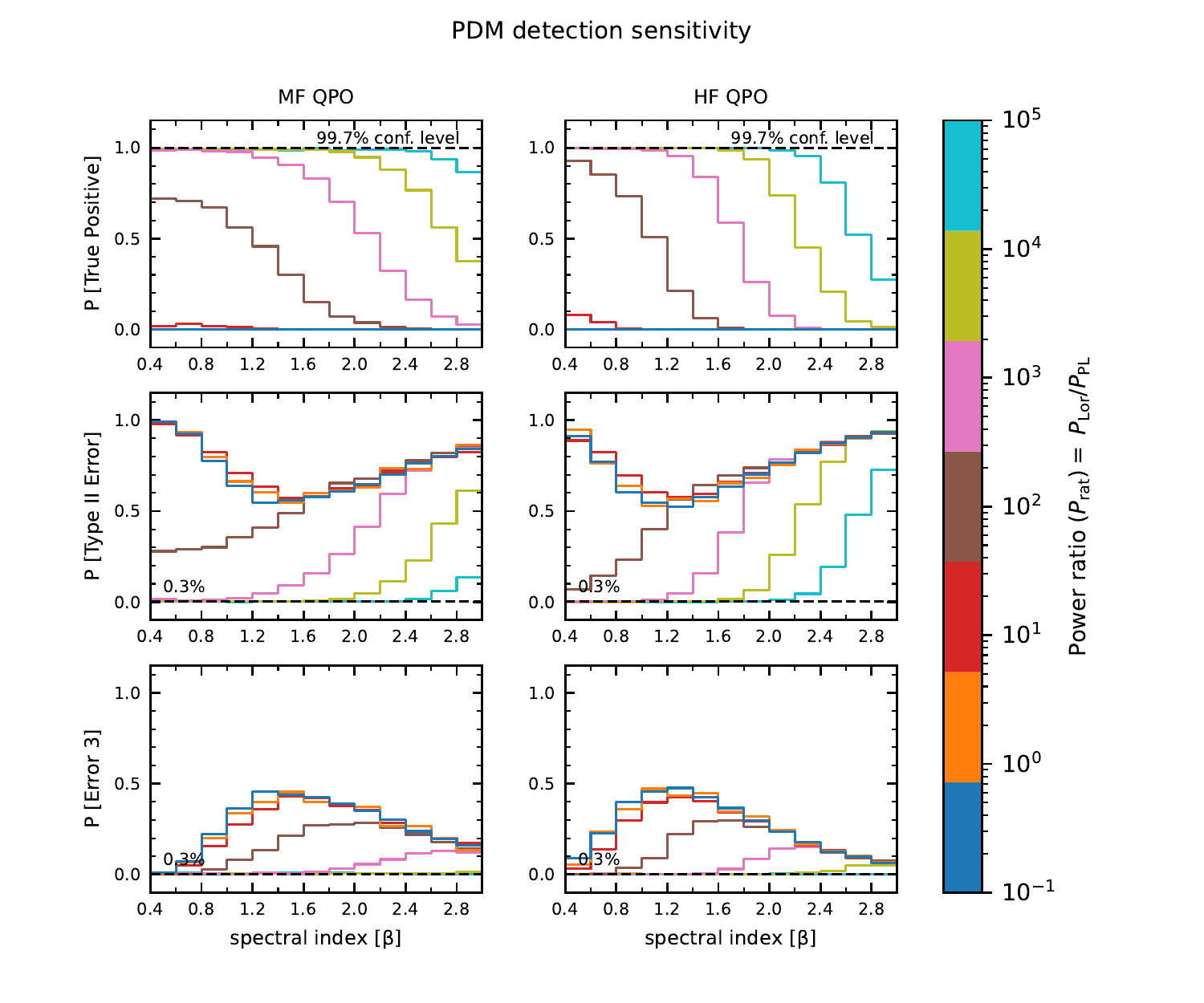}

\caption{Top panel: The probability of detecting the first harmonics of the periodic
          signal in the expected frequency range, Middle panel: The probability of detecting the first harmonics of the quasi-periodic
          signal in the wrong frequency range, Bottom panel: The probability of detecting false negative of the first harmonics of the quasi-periodic
          signal all as a function of power ratio $P_{\rm rat}$ and
          red noise PSD power-law slope $\beta$ for the MF \& HF QPO using PDM. }
\label{fig:QRNPDM}
\end{figure*}

\subsubsection{False Alarm Probabilities}
\textit{Type I error (false positive): We calculate how often H1 is incorrectly inferred true when H0 is intrinsically true} from the 1000 MCS of pure red noise light curves generated for a range of spectral indices. We determine the rate at which a pure red noise light curve can minimize the test statistic to low values at ANY frequency, with at least one occurence per realization and can be falsely registered as an QPO signal.

We register the number of times $\theta_{\rm min}$ for a given PDM drops below the 99.9 per cent upper confidence limit of $\theta_{\rm min}$ obtained from the ``ideal'' (Lorentzian-only) case of HF QPO having the highest value of 0.88, for each realization of the red noise light curves, to determine the probability of the false positives. In the case of unbroken power-law model as shown in Fig.~\ref{fig:pdmfap}, we see that the rate of false positives is $>0.3$ per cent at all the tested $\beta$ values. When we neglect the lower frequency bins corresponding to timescale $>$ $1/3$ duration, even though the rate of false positive is still $>0.3$ per cent at all slopes, we see that there is a decrease in the rate of false positives especially towards the higher spectral index slopes. We also see a slight increase in the rate of false positives at spectral index slopes $0.6 \lesssim \beta \lesssim 1.4$ because around these spectral index slope values, the red noise has $0.65<\theta_{\rm min}<$ 0.8 at frequency bins corresponding to timescales $\sim 1/3$--$1/9$ duration. In the case of broken power-law model, the rate of false positives is $>0.3$ per cent for all the tested combinations of $\beta =1.0$, 2.0, 3.0 breaking to $\gamma=0.0$, 1.0, 2.0 towards the lower frequency after the break at the mid-temporal frequency before and after excluding the lower frequency bins.

It is important to note that after neglecting the lower bins, less than $0.3$ per cent fractions of red noise simulations having unbroken and broken power-law model can attain a value of $\theta_{\rm min} \lesssim 0.65$ and 0.5 on using the PDM for all the tested values of slopes respectively.

\subsection{The behavior of PDM for QPOs mixed with red noise}

Once again, we perform MCS for the sum of a $Q=30$ Lorentzian and an unbroken power-law red noise continuum, with slopes spanning 0.4--3.0 in steps of ${\Delta}\beta=0.2$, and power ratios log($P_{\rm rat}$) ranging from $-$1 to +5 in steps of ${\Delta}$log($P_{\rm rat})=1$.

\textit{True-positive detections:} We register true positive detections when $\rm \theta_{\rm min}$ falls into the 99.9 per cent confidence range in $\theta$--frequency space as determined from the ``ideal'' (Lorentzian-only) scenarios above. For example, in Fig.~\ref{fig:pdmerr}, we plot the PDM periodogram for few realizations of lightcurve having a MF QPO signal mixed with red noise of $\beta =2.6$. The black points are the lowest minimum ($\theta_{\rm min}$) for a given PDM periodogram and the green shaded region is the $99.9$ per cent contour limits of the ``ideal signal'' in the $\theta$--frequency space expected for the MF QPO signal. Hence, only those points which falls into this region will be registered as true positives.

We find that in order to ensure that the probability of correctly registering true positives with $\sim$99.7 per cent, one always requires an extremely high value of log($P_{\rm rat}$): typically $>5$ or more. We see that in the case of MF QPO, the signal is not detectable with the 99.7 per cent significance. As shown in Fig.~\ref{fig:QRNPDM} for the MF QPO, values of log($P_{\rm rat}) \sim$ 4, 5 only approach reliable detections of about $99$ per cent when $\beta \lesssim 1.8$, 2.4 respectively of which $>$85--90 per cent fraction of the simulations has $\theta_{\rm min} < 0.65$. Finally, for the HF QPO, values of log($P_{\rm rat}$) $\sim$ 4, 5 has 99.7 per cent significance detections when $\beta \lesssim$ 1.6, 2.0 respectively and log($P_{\rm rat}$) $\sim 3$ only approach reliable detections of about $99$ per cent when $\beta \lesssim 1.0$. In the case of HF QPO signals mixed with red noise $>$ 30--40 per cent fraction of the simulations with reliable detections has $\theta_{\rm min} < 0.65$. 

The above analysis was performed considering values of $\theta_{\rm min}$ occurring at any frequency while excluding all frequencies less than a frequency of $\sim$ 0.012 d$^{-1}$, which corresponds to one-third of the duration. We also repeated the analysis including the lowest-frequency bins, but this action did not yield significant changes to the results of significant detection of the period. In fact, excluding the lowest bins helps only in reducing the detection inaccuracy error of registering $\theta_{\rm min}$ at the wrong frequency (i.e, towards the lower bins) which will be explained in $\S4.3.1$.

\subsubsection{False negatives and detection inaccuracy errors}

\subsubsection*{Type II Error (false negative): H1 is inferred false incorrectly when a QPO is intrinsically present.} When the value of $\theta_{\rm min}$ does not fall below the upper limit threshold determined from the ``ideal'' case (regardless of frequency), we register that trial as a false negative. For example, in Fig.~\ref{fig:pdmerr}, some of the black points which represent the lowest minimum ($\theta_{\rm min}$) for a given PDM periodogram obtained for light curves having MF QPO mixed with red noise, lies above the upper limit on $\theta$ (0.805) below which MF QPO signals are expected. Hence, they will be registered as false negative.

As shown in Fig.~\ref{fig:QRNPDM}, for the MF QPO, the type II error rate is below $\sim 0.3$ per cent for values of log($P_{\rm rat}$) $\sim$ 5, or 4 when $\beta \lesssim 2.4$, or 2.0 respectively and for the HF QPO, the false negative rate is $\lesssim 0.3$ per cent for values of log($P_{\rm rat}$) $\sim$ 5, 4, 3 when $\beta \lesssim 2.0$, 1.6, 1.0 respectively. We see that with decreasing log($P_{\rm rat}$) and/or increasing steepness $\beta$, the rate of the false negatives increases. We note that at log($P_{\rm rat}) \lesssim 2$, the rate of false negatives increases towards the lower spectral slopes of $\beta \lesssim 1.4$. This is because at the lower power values of QPO against the red noise, the probability of finding a minimum $\theta$ at the expected contours for the MF \& HF signals decreases which in turn increases the rate of false negatives around these region.

\subsubsection*{Detection inaccuracy error (Error 3): H1 is intrinsically true and also registered as true, but the QPO is detected at the wrong frequency.} We register detection inaccuracy errors when $\theta_{\rm min}$ falls below the upper limits derived in the ``ideal'' case, but also falls outside the expected frequency bounds. For example, in Fig.~\ref{fig:pdmerr}, the black points which are the lowest minimum ($\theta_{\rm min}$) for a given PDM periodogram obtained for light curves having MF QPO mixed with red noise, lies below the upper limit on $\theta$ (0.805), but does not lie within the expected frequency range for MF QPO. Therefore, when we have $\theta_{\rm min}$ of the PDM periodogram occuring in the yellow shaded region, then it is registered as Error 3.

When one blindly searches for $\theta_{\rm min}$ across all frequencies, including in particular the lowest-frequency bins, the probability of encountering a detection inaccuracy error is generally very high, and it increases towards higher values of $\beta$ and/or lower values of log($P_{\rm rat}$). Now, when we neglect these lower frequency bins, the rate of finding the $\theta_{\rm min}$ at the wrong frequency bins decreases both along the log($P_{\rm rat}$) and the spectral index slope values. In Fig.~\ref{fig:QRNPDM}, we see that even though the rate of detection inaccuracy error is $>0.3$ per cent, it is still significantly low ($\lesssim 1$ per cent) at log($P_{\rm rat})\sim$ 4--5, for $\beta \lesssim 2.6$ for both the MF \& HF QPO.

%\section{REDUCING THE EFFECT OF RED NOISE PROCESSES}
%\input{filteringv4.tex}
\section{REALISTIC UNEVEN SAMPLING PATTERNS}
\begin{figure*}
	\includegraphics[width=\linewidth]{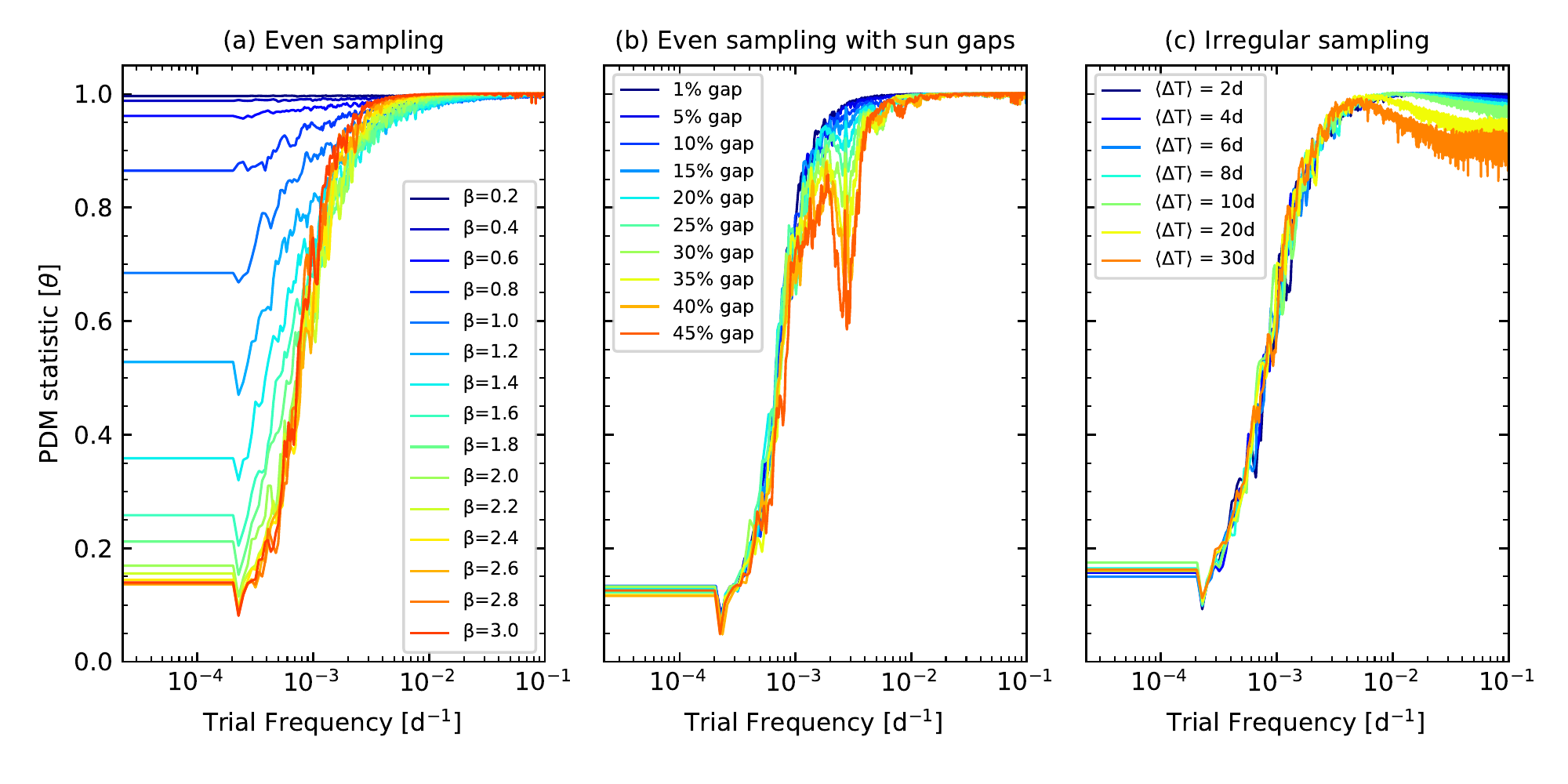}
    \caption{The distribution of the 99.9 per cent lower limit of the PDM statistic value $\rm \theta$ at each test frequency of pure red noise 10 year long LSST type light curve for (a) Evenly sampled data at different spectral index slopes, (b) Evenly sampled data with yearly sun gaps of 45 per cent data gap for $\rm \beta= 2.0$, (c) Irregularly sampled data having different average sampling rate for red noise slope $\rm \beta= 2.0$. }
    \label{fig:PDMRNsamp1}
\end{figure*}

The AGN light curves that we get from observations are usually unevenly sampled, and typically contain data gaps and/or irregular sampling. It is thus important to explore the effects of sampling patterns when using the ACF and PDM for searching for periodic signals. It is impossible to explore all types of potential sampling patterns, so here we choose representative sampling patterns that could plausibly be observed by ground-based optical large-area surveys such as LSST, PanSTARRS, ATLAS, etc. --- continuous monitoring quasi-regularly, but interrupted by yearly sun-gaps of a few months --- or ground-based radio surveys such as OVRO --- continuous monitoring usually year-round, with no major gaps, but with somewhat irregular sampling. We check how irregular sampling and the presence of regular gaps such as yearly sun gaps impact the signatures of pure red noise processes and impact detectability of a QPO when it is mixed with red noise.

We conduct Monte carlo simulations of 10 year long light curves similar to a generic optical survey such as LSST or ATLAS, first with evenly sampled data with and without different yearly sun gaps, and then we test different irregular sampling patterns, again with and without the yearly sun gaps. We perform this analysis only for a subset of the parameters explored above, as this section is intended as an exploration of the effects of only basic facets of irregular data sampling. Particularly for more complex data sampling patterns, readers are strongly encouraged to conduct their own simulations following our work as an example.

\subsection{PDM test for Lorentzian profile}

We repeat the analysis as in $\S$4, where we first consider the ideal case of QPOs only, represented by simple Lorentzians of quality factor $Q=30$. We start with evenly sampled data, assuming 10-year long light curves having a sampling rate of one point every day. In lieu of testing potential QPOs across all frequencies between 1/(3650 days) and the Nyquest frequency of 1/(2 days), We choose three representative test frequencies $f_{\rm L}$: 1.3, 5.77, and 22.04 $\times10^{-8}$ Hz, corresponding to timescales of 2.3, 0.55, and 0.144 years, and spanning 4.3, 18.2, and 69.5 cycles, respectively; we refer to them low-, medium- and high-frequency (LF, MF, and HF) QPOs.

We first simulate light curves to determine the region of frequency--$\theta$ space where the ``ideal'' signal occurs. We determine the 99.9 per cent confidence limits on $\theta_{\rm min}$ (upper limits) : 0.705, 0.769, and 0.864 for the LF, MF and HF QPOs respectively; $\nu_{\theta_{\rm min}}$ (lower, upper bounds) are (0.0005, 0.0014), (0.0045, 0.0055), and (0.0181, 0.0204) for the LF, MF and HF QPO respectively. These values delineate the ranges within which to search for signatures of 
QPOs when we consider mixtures of a QPO and red noise. 

Similarly, we determine the 99.9 per cent confidence limits on $\theta_{\rm min}$ and $\nu_{\theta_{\rm min}}$ first for the unevenly sampled data having different average sampling rates of 2 -- 10 days in steps on 2 days, 20 \& 30 days; and then for evenly sampled data \& irregularly sampled data (avg. sampling rate of 8 days) having sun gaps of different values (36.5 d, 0.5 -- 4.5 yrs in steps of 0.5 yrs): We see that the the range of frequency--$\theta$ limits are about the same ---
with only $\sim$4--6 per cent deviation --- as the values for evenly sampled data.

\subsection{The behavior of the PDM for pure red noise processes}

We perform MCS for each of the different sampling patterns of red noise process of unbroken power law PSD model and see the effect in the PDM compared to the evenly sampled data.

\textit{Even sampling pattern:} We again generate pure red noise light curves using an unbroken power-law PSD model for a wide range of slopes $\beta$ (0.2--3.0), in steps of ${\Delta}\beta$=0.2, simulating 1000 light curves at each step for the evenly sampled data for 10 years duration having one point per day, and measure their PDMs.

The resulting 99.9 per cent confidence lower limits on $\theta$ at each trial frequency are plotted in Fig.~\ref{fig:PDMRNsamp1}. In the figure, we see that the PDM looks similar for the evenly sampled data as in $\S$4.2.1 where $\theta$ reaches very low values. We notice that at time scales about 1/3rd--1/4th of the duration of the light curve, there is a turnover in the values of the statistic minimum value where it starts to get below 0.85--0.8 and with increasing time scale. The statistic minimum can reach lower than 0.5 at much longer timescales ($\lesssim$ one-half the duration) and can get even as low as 0.2 when PSD slopes are $\gtrsim 1.8$.

\textit{Even sampling with yearly sun gaps:} Now we test the effects of yearly sun gaps in the evenly sampled light curves in the PDM. So we produce 1000 light curves of unbroken power-law PSD model having slope of $\beta = 2.0$ for each different range of gaps from 1 per cent of the duration corresponding to 0.1 yrs and slowly increasing the amount of gap in steps of 5 per cent of the duration corresponding to a range of 0.5--4.5 years of gap in the 10 year long data for demonstration on the effects of sun gaps. In Fig.~\ref{fig:PDMRNsamp1}, we plot the 99.9 per cent confidence lower limits on $\theta$ at each trial frequency for each different gaps. We see that as the amount of gap in the data increases, it produces an extra feature, where there is a sudden dip in the $\rm \theta_{min}$ which can get lower than 0.7 at frequency $\sim$ 1/(365d), when the gaps are greater than 35 per cent of your complete data apart from the deep minimum observed towards the lowest frequencies ($\lesssim$ 1/3 of the duration), which are not significantly affected by the sun gaps. 

Again, to see the effect of different slopes of red noise with data gap in the PDM, we produce red noise light curves using an unbroken power-law PSD model for a wide range of slopes $\beta$ (0.4--3.0), in steps of ${\Delta}\beta$=0.2, having 45 per cent gap in the data, simulating 100 light curves at each step and find the 99.9 per cent confidence lower limits on $\theta$ at each trial frequency for the different spectral slopes. The dip at a frequency of 1/(365d) becomes stronger for relatively steeper PSD slopes, including falling below $\sim$0.8 for values of $\beta$ steeper than $\sim$ 1.2. 

We thus recommend to avoid the frequency space corresponding to lower than roughly 1/3--1/4 of the duration of the lightcurve, as well as other obvious timescales range corresponding to the period of the gaps, as these timescales can see spurious deep minimum features in the PDM.

\begin{figure}
	\includegraphics[width=\linewidth]{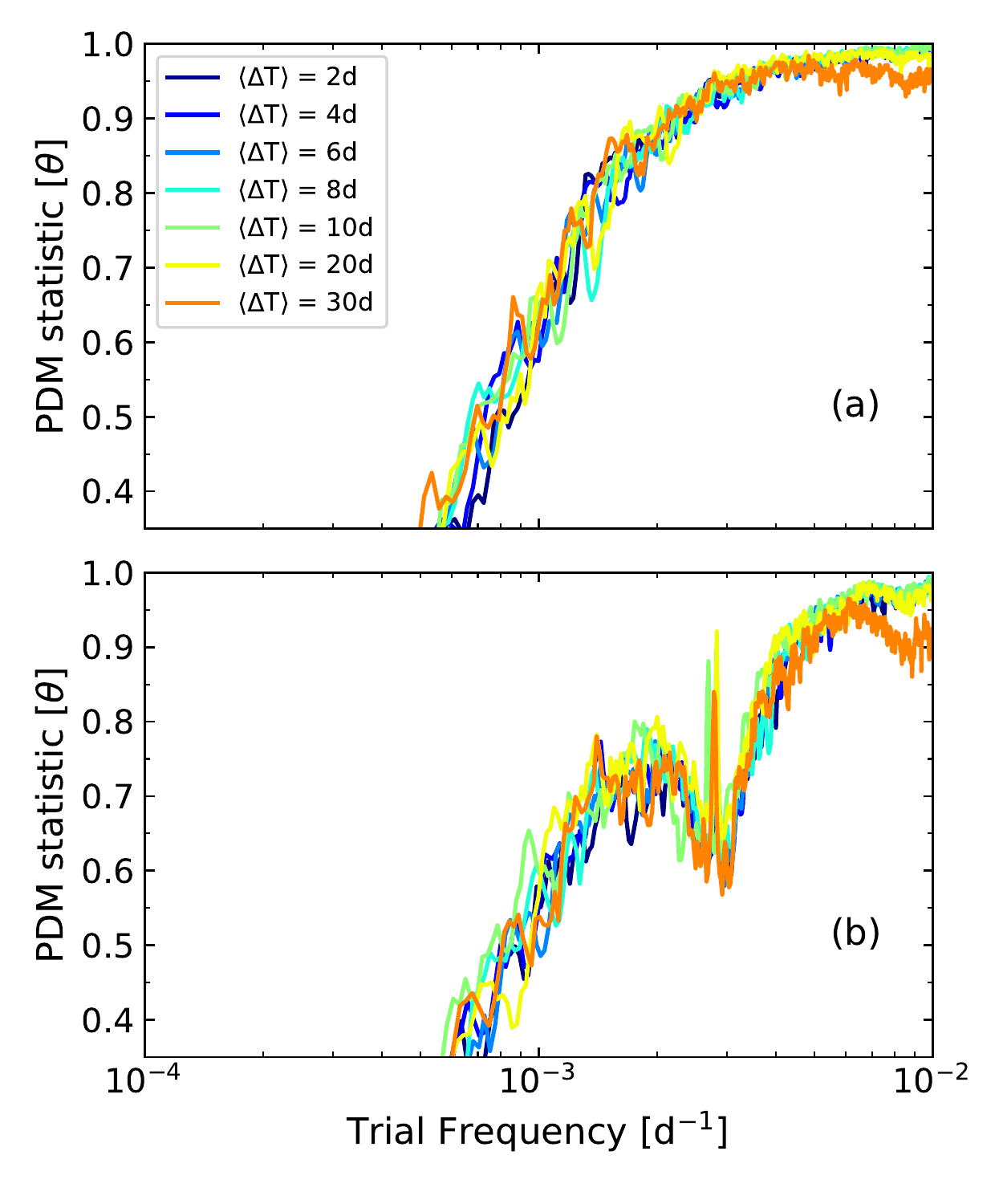}
    \caption{The distribution of the 99.9 per cent lower limit of the PDM statistic value $\rm \theta$ at each test frequency for a randomly sampled pure red noise light curve having spectral index slope $\beta \sim 2.0$ with (a) 10 per cent yearly gap and (b) 45 per cent yearly gap for different average sampling rates.}
    \label{fig:PDMRNsamp2}
\end{figure}

\textit{Irregular sampling:} Here we test the effects of irregular sampling without any large data gaps. We again simulate 10-year long light curves with daily sampling, but we resample by randomly selecting points every $\rm {\Delta} T$ days, where we explore average values of $\rm \langle {\Delta}T \rangle =$ 2, 4, 6, 8, 10, 20, and 30 days. We assume $\beta = 2.0$, and we again produce 100 light curves and determine the 99.9 per cent confidence lower limits on $\theta$ at each trial frequency; the results are plotted in Fig.~\ref{fig:PDMRNsamp1}. Here, it is the relatively higher temporal frequencies that are affected: particulary for $\rm \langle {\Delta}T \rangle \sim$ 20--30 days, $\rm \theta_{min}$ consistently departs from just under unity to roughly $0.95$--$0.87$ for frequencies corresponding to timescales shorter than roughly $1/20$ of the duration. However, no narrow-band artefacts are induced.

\textit{Irregular sampling with yearly sun gaps:} We test the effects in the PDM for the combination of irregular sampling and yearly data gaps. For an input unbroken power-law PSD model $\beta=2.0$, we determine the 99.9 per cent confidence lower limits on $\theta$ at each trial frequency for each of the different avaerage sampling rate having 10 per cent wide data gaps and then 45 per cent wide data gaps (the latter is not uncommon for many ground-based optical programmes for targets near the equator). As shown in Fig.~\ref{fig:PDMRNsamp2}, for the 10 per cent wide gap case with irregular sampling there is not much difference in the PDM compared to when there are no data gaps. For 45 per cent wide data gaps, we again see the spurious dip near timescales of 1/(365d) as before.

\subsection{The behavior of PDMs for QPOs mixed with red noise}
We perform MCS for the sum of a $Q=30$ Lorentzian and an unbroken power-law red noise continuum, with slopes spanning 0.4--3.0 in steps of ${\Delta}\beta=0.2$, and power ratios log($P_{\rm rat}$) ranging from $-$1 to +5 in steps of ${\Delta}$log($P_{\rm rat})=1$, now for the 10 year long 
light curves for each of the different sampling patterns. For brevity we test only the MF QPO,
though results were qualitatively similar for the HF QPOs.

We test for evenly sampled data for the 10 year long light curve having one point per day to check for the detection probability along the different spectral index slopes and along the different power ratios. We find that the MFQPO has a detection sensitivity range of significant detection of 99.7 per cent only at log($P_{\rm rat}) \gtrsim 5$.      

We repeat the test for different uneven sampling patters (a) Even sampling with yearly sun gaps of 40 per cent, (b) Random sampling pattern having an average sampling rate of 8 days and (c) Random sampling having an average sampling rate of 8 days  with yearly sun gaps of 40 per cent for the 10 year long light curve to check for the detection probability along the different spectral index slopes and along the different power ratios. We see that the detection sensitivity for the different sampling patterns to be the same as when it was evenly sampled and thus we conclude that for the PDM, sampling artefacts do not play a signifcant role in true-positive QPO detection densitivity.     
\section{DISCUSSION}
Through extensive Monte Carlo simulations of quasi-periodic processes and mixtures of quasi-periodic and red noise processes, we empirically explored the regions of parameter space where a (true positive) detection of a QPO with the ACF or PDM is statistically robust.

We also empirically explored the signatures produced by pure red noise processes when using the ACF and the PDM, and quantified the regions of parameter space where false positives are likely to be encountered.

\begin{figure}

\includegraphics[width=\linewidth]{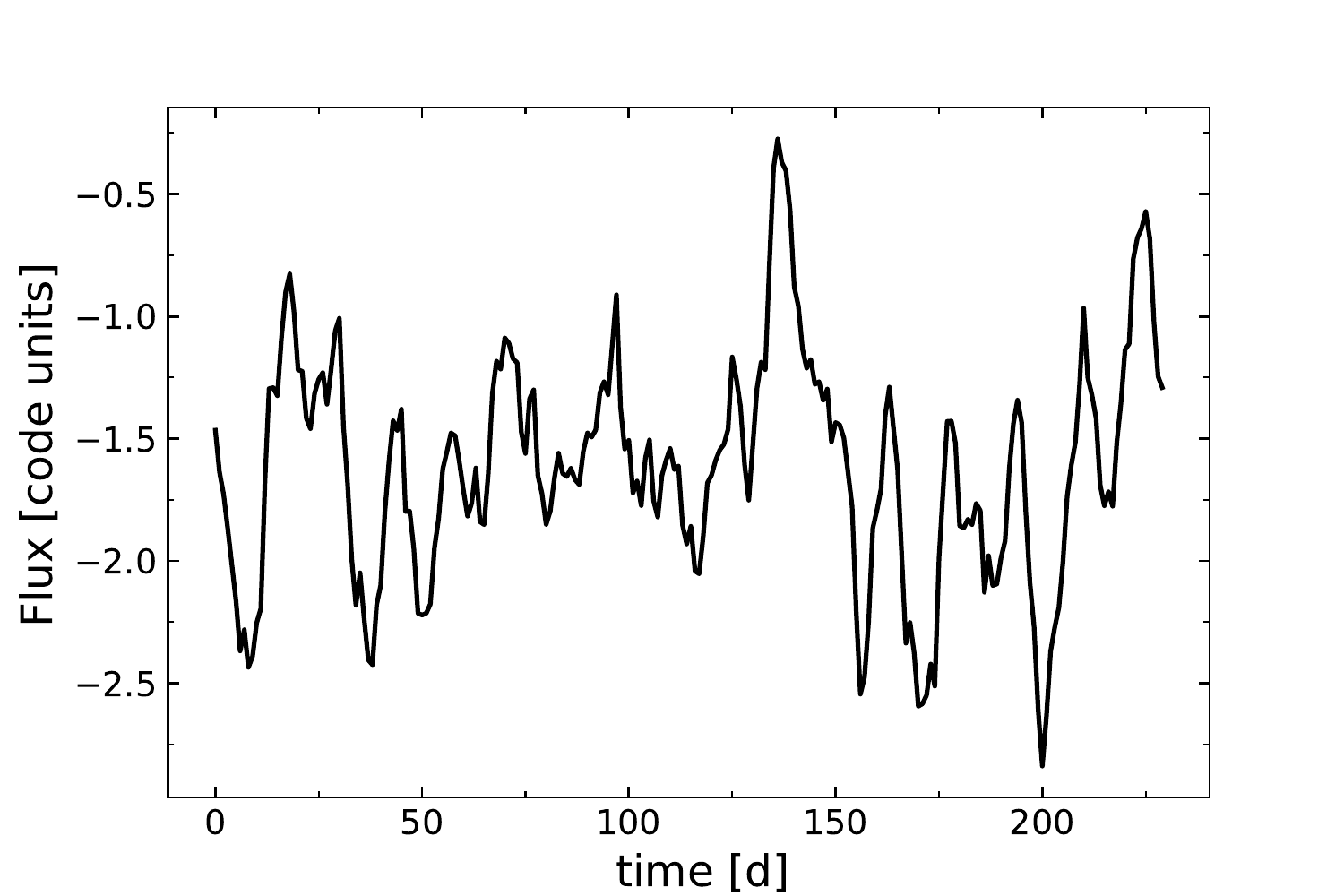}
\caption{A sample light curve corresponding to power spectra of MF QPO signal of log($P_{\rm rat})=2$ against the red noise continuum having $\beta = 1.8$, where mild features of quasi-periodicity are visually evident in the light curve. However, such a light curve would very likely ( likelihood) yield no detection with an ACF or PDM. }
\label{fig:demolc}
\end{figure}

The light curve sampling patterns we explored are admittedly limited in number but relevant to long-term (multi-year) observing programs with sustained monitoring, and with sampling patterns relatively poorer than that required for a DFT.  With these sampling patterns, we
generally find that:

\begin{itemize}

\item In studying the effect of pure-red noise processes in the ACF, assuming an unbroken power-law PSD model and evenly sampled data, we see that pure red noise can produce false positives determined across the time lag at a rate more than $0.3$ per cent for all tested values of the spectral index slope, which is significantly high to be neglected. In fact, correlation coefficients of the first peak can easily reach above 0.5--0.7 at $\beta \gtrsim 1.4$, and can reach even higher values ($\sim$ 0.9) at steeper slopes of $\beta \gtrsim 2.2$. The second peak after the zero lag determined across the time lags independently can also have a correlation value $\sim$ 0.7 at $\beta \gtrsim 1.0$ and are easily misintepreted as an QPO signal at these corresponding time lags.

One can robustly avoid peaks with correlation coefficients above 0.45--0.5 caused by pure red noise by restricting oneself to lags less than roughly 1/3 of the light curve duration and thus have the false positive rate across the time lags to be $<0.3\%$ while searching for a QPO signal mixed with red noise. A broken power-law input PSD (e.g., breaking from slopes from $\beta =2.0$ to $\gamma= 1.0$ or to 0.0) yield qualitiatively similar results.

\item In the analysis of pure red noise processes in the PDM, assuming an unbroken power-law PSD model and evenly sampled data, we see that towards the lower frequency bins --- corresponding to timescales longer than roughly one-third of the duration--- $\rm \theta$ consistently dips below 0.8--0.85. In fact, about 99.7 per cent of the pure red noise simulations with PSD slope of $\beta \gtrsim 2.4$ can yield values of $\rm \theta_{min}$ below 0.6. Therefore, it is to be noted that low values of $\theta$ in a given PDM do not automatically mean that a QPO is present e.g., at low frequencies.

After excluding the lowest bin frequencies (corresponding to longer than $1/3$ of the duration), we see that the fraction of pure red noise-only trials that can attain a value of $\theta_{\rm min} < 0.6$ is less than 1 per cent for all the tested combinations of broken and unbroken power-law PSD models. \textit{Hence, merely neglecting the lower frequency bins in the PDM is a straightforward way to minimize false postitive detections.}
  
\item We considered mixtures of red noise processes and a $Q=30$ QPO, and explored detection sensitivity ranges in both the ACF and PDM for a range of red noise slopes and QPO strengths relative to the red noise (quantified by log($P_{\rm rat}$)), neglecting timescales $>1/3$ duration, we find:

\begin{itemize}

\item Detection of the period with statistical significance ($\gtrsim 99.7\%$) depends strongly on both the strength of the QPO against the red noise and the steepness of the red noise PSD slope. In general, the probability of detection of the signal at the correct period gets lower with increasing steepness of the red noise slope and decreasing power of the QPO against the red noise.

\item Particularly for relatively steep PSD slopes, extremely large values of log($P_{\rm rat}$) are required for robust detections of a QPO in both the ACF and PDM. 

\item For illustration, In Fig.~\ref{fig:demolc}, we have the light curve corresponding to power spectra of MF QPO signal of log($P_{\rm rat})=2$ against the red noise continuum having $\beta = 1.8$. Even if mild features of quasi-periodicity are evident in the light curves, these power values of QPO would likely not register a true positive detection in either the ACF (52.3 per cent likelihood of not registering a true positive detection) or PDM (86 per cent likelihood of not registering a true positive detection) at any time scale.

\item With the ACF, we generally find that QPOs are detected robustly only when log($P_{\rm rat}) \gtrsim 5$ for $\beta \gtrsim 2.4/2.2$ for a MF/HF QPO, respectively. Detections at values of $\rm log(P_{rat}) \sim 3$ are attainable only at lower spectral index slopes, $\beta \lesssim 1.2$.
 
\item For the PDM, we require $\rm log(P_{rat}) \sim $ 4, 5 for $\beta \lesssim 1.8$, 2.4 even for a 99 per cent detection for MF QPO. Detections at values of $\rm log(P_{rat}) \sim 4, 5$ are possible for $\beta \lesssim 1.6$, 2.0 for the HF QPO.

\end{itemize}

\item We explore the effects of regular data gaps and irregular sampling in the PDM. Irregular data sampling causes $\rm \theta_{min}$ to become low ($\sim$0.95) across broad swaths of frequencies, but true-positive detection rates for mixtures of QPOs with red noise are not significantly impacted. When a regular data gap such as a yearly sun gap are included, it produces a relatively narrow-band feature that is confined to e.g. 1/(365d).
  
\end{itemize}

When there is a claim of a QPO signal from AGN in the literature, then one should be able to not just articulate its frequency \& RMS, but also the form of the underlying red noise continuum and the uncertainities related to these parameters. In the case of ACF/PDM, it is not easy to determine the underlying PSD shape. We conclude that since the ACF/PDM tend to produce false alarm rates greater than $0.3\%$ and given the general quality of AGN data gathered so far, it is advisable that the community should be cautious and refrain from publishing claims of QPO using the ACF/PDM (until they have a reliable and significant detection considering proper null hypothesis), since the detection rate is generally low and not significant until one reaches very high values of power of QPO against the red noise continuum.

In $\S$6.1, we briefly discuss some of the main problems affecting many of the claims of periodicities in the literature in AGN. In short, they include 1) sampling too few cycles of a ``signal'', as red noise can produce spurious sinusoid-like trends on the longest timescales, and/or 2) mistreatment of the null hypothesis: white noise, or red noise with insufficiently steep power-law slope, is assumed; the likelihood of false positives tends to increase towards steeper PSD slopes.

Another issue regarding claims in the literature, though, is that there is rarely any mention of the inferred RMS or RMS/mean (hereafter referred to simply as RMS for simplicity) of the claimed periodic signal.  While the value of the period of course encodes physical info about the variability mechanism, so does the RMS of the QPO; if parameter constraints from other analyses are known, a consistency check can further help test if the claimed QPO is real. We discuss this topic further, with a couple brief applications, in $\S$6.2.

\subsection{Some of the claims in the literature and basic pitfalls regarding ACF and PDM usage}

One major caveat in searching for QPOs against a red-noise background, as pointed out by e.g. \citet{2016MNRAS.461.3145V}, is that pure red noise processes, particularly those with relatively steep PSD slopes ($\beta > \sim 2$), can produce quasi-sinusoid-like, ``W-shaped'' segments of light curves for three or four ``cycles.'' 

For observations of a finite length of a red noise process, the dominant trends on timescales longer than (very roughly) one-third of the full duration may be quasi-sinusoidal. These trends can potentially ``trick'' any period-searching method --- ACF, PDM, fitting sinusoidal functions to light curves --- into falsely identifying a period, without a proper calibration against a null hypothesis in place (e.g. \citealt{1978ComAp...7..103P}). 

In this paper, we empirically quantify how pure red noise processes can produce false signatures in both the ACF and the PDM reminiscent of those produced by a pure Lorentzian signal --- strong correlation peaks in the ACF, and dips in $\theta_{\rm min}$ in the PDM --- particularly on timescales longer than roughly one third of the light curve duration.

The problem of observing too-few cycles can potentially be resolved by extending observations. For example, \citet{2015Natur.518...74G} claimed a period of $\sim$1900 days in an optical light curve of the quasar PG~1302--102 by sinusoidal fitting of a $\sim$10-year light curve; \citet{2016MNRAS.461.3145V} demonstrated the light curve's consistency with pure red noise. Moreover, an additional five years' monitoring reported by \citet{2018ApJ...859L..12L} failed to confirm this period.

A second major caveat is that developing and testing a proper null hypothesis (pure red noise, no QPO) is critical for testing the statistical significance of a candidate period (especially in the absence of having the luxury of being able to extend monitoring). Testing against a null hypothesis consisting of Poisson (instrumental) noise only is clearly insufficient; some papers such as \citet{Li_2009} do not attempt any such significance test. We empirically see that with both the ACF and PDM, the likelihood of encountering a false positive from a pure red noise process is extremely sensitive to the value of the underlying power-law PSD slope, and false positives can be significant in number particularly for $\beta > \sim 2$.  We recognize that it can be difficult to measure a periodogram and obtain an unbiased fit when data sampling is poor, but having at least a roughly accurate idea of $\beta$ is thus critical for being able to form a proper null hypothesis.

Below we briefly discuss these warnings in the context of using the ACF and PDM, and point out some examples in the literature of misinterpretation.  Our intention is not to single out individual authors; our intention is to simply focus on discussing common mistakes, citing a few select examples in the literature only.
A caveat is that data sampling in these cited publications are always not exactly the same as in our simulations in terms of dynamic range of frequencies explored, frequency resolution, number and size of data gaps, and the irregularity of data sampling. In addition, estimates of power-law slope $\beta$ are usually difficult to know reliably, sometimes because data were too sparse for a reliable periodogram fit.
However, we believe that differences in sampling should not significantly impact our qualitative conclusions about these works. Moreover, these works each have data sampling such that the ratios of maximum to minimum temporal frequencies probed is usually of order a few hundred, roughly similar to what we explored in this paper.

\subsubsection{General issues with ACF, with a few select examples:}
One basic problem with attempting to use the ACF for period-searching, as demonstrated in $\S$3.2, is that pure red noise processes can produce spurious bumps and wiggles at all timescales (particularly relatively longer timescales and when $\beta > \sim 1.5$) which might be mistaken for a QPO signature.

ACFs in citations claiming a QPO frequently do not show the clear cosine-like behavior expected when a QPO dominates; both \citet{Li_2009} and \citet{2002A&A...381....1F}, for instance, simply identify multiple local peaks in their ACFs as candidate signals, but these are likely artefacts of red noise, e.g. with $\beta \sim 1-2$.

\citet{2001A&A...377..396R}, \citet{2011JApA...32...79L} and \citet{2021MNRAS.501.1100R} each claim QPO signatures in peaks occuring at timescales $>$ 1/3 -- $\sim$ 1/2 of the light curve durations, with values of $r_{\rm corr}$ peaking at $\sim$ 0.5--0.7.
However, we demonstrated that for both unbroken power-law PSDs with $\beta > 1.4$ and broken power-law PSDs, $r_{\rm corr}$ can frequently peak above 0.7 (and $r_{\rm corr}$ can even frequently peak above 0.9 when $\beta > \sim 2.2$ very easily, especially at timescales $>$ 1/3 of the duration.
For \citet{2011JApA...32...79L}, the fundamental issue is that their light curve samples 2.5 cycles of a sinusoid-like trend ("W-shape") and is consistent with red noise likely corresponding to a very steep PSD slope, probably $\beta \sim 2$ or $3$.

\subsubsection{General issues with PDM, with a few select examples:}

One basic issue with attempting to use the PDM for period searching was demonstrated in $\S$4.2: pure red-noise processes yield spuriously low values of $\theta_{\rm min}$ at relatively low frequencies / long timescales e.g. $> \sim$ 1/3 of the light curve duration. For example, even in the limit of perfect data sampling, red noise processes across all range of $\beta$ tested can cause $\theta_{\rm min}$ to drop to below values of typically e.g. $\sim$0.80--0.85 below a frequency corresponding to $\sim30$ per cent of the duration for the sampling patterns and red noise slopes simulated in this paper.

In the literature, claims of periods hinge on observed localized minima in the PDM, but these minima tend to occur towards long timescales, $>$ $\sim$ 1/3--1/4 of the duration: \citet{2002A&A...381....1F}, \citet{2011JApA...32...79L}, and \citet{2020ApJ...896..134P}.

Note that \citet{2011JApA...32...79L} and \citet{2002A&A...381....1F} have used the \citet{1971Ap&SS..13..154J} method, which is numerically related to the PDM: both calculate the sample variance within each test phase bin and consider the sum of those variances. However, the PDM is normalized by the overall variance in the light curve. Meanwhile, \citet{2002A&A...381....1F} and \citet{Li_2009} used a renormalized Jurkevich parameter $V^2_{\rm m}$ following \citet{1992A&A...264...32K}, defining a fraction $f \equiv (1 - V^2_{\rm m})/V^2_{\rm m}$, but it does not take into account the number of degrees of freedom; the expectation value in the absence of any periodicity consequently can differ significantly from 1.
In this paper, we do not simulate the Jurkevich method specifically, and a direct comparison of our values of $\theta_{\rm min}$ to values of $V_{\rm m}^2$ as a function of timescale is not immediately straightforward. Nonetheless, given the characteristics of pure red noise processes, claims of periods occurring at long timescales using either method should be relatively suspect.

\subsection{Application to detection of periods for selected physical situations}

\begin{table*}
  \begin{tabular}{|l|c|c|c|c|}
    \hline
    \hline
%    \multirow{2}{*}{} &   %% 
      &  \multicolumn{2}{c}{Mass ratio $q=0.05$} &
         \multicolumn{2}{c}{Mass ratio $q=0.5$} \\

    $N_{\rm E}$ & NUV & V & NUV & V  \\
    \hline
    0.05 & 0.752 (5.67) & 0.471 (5.27) & 0.784 (5.70) & 0.504 (5.33) \\
%    \hline
     0.1 & 0.471 (5.27) & 0.281 (4.82) & 0.504 (5.33) & 0.307 (4.89) \\
%    \hline
     0.5 & 0.095 (3.87) & 0.074 (3.66) & 0.105 (3.96) & 0.082 (3.75) \\
%    \hline
     1.0 & 0.030 (2.88) & 0.030 (2.88) & 0.033 (2.96) & 0.033 (2.96) \\
    \hline
    \hline
  \end{tabular}
  \caption{The values of RMS/mean estimated from the light curves of
    highly-inclined SMBH binary systems, based on the flare emission
    profiles in DD18. Listed here are values for two selected values
    of mass ratio, four selected values of $N_{\rm E}$, the fraction
    of an Einstein radius separating the lens and the source at
    closest approach, and two wavebands, NUV and V. Also listed, in parentheses, are
    estimates of log($P_{\rm rat}$) assuming a mixture of a $Q=30$ QPO at the MF QPO frequency, $8.0\times10^{-7}$ Hz, and a
    continuum PSD with $\beta=2.3$ and a normalization of 20 Hz$^{-1}$
    at $10^{-6}$ Hz.}
  \label{tab:rmswithlogpr}
\end{table*} 
When there exists a periodic signal mixed with red noise in the emission from a source, and if the red noise power spectral characteristics can be estimated, then, for a given sampling pattern and a given test frequency, the threshold values of power ratio required for a successful detection of a periodic signal with statistical significance using PDM or ACF can be translated into threshold values of RMS for the periodic component.

Such threshold values can be used to define the regions of parameter space for a given physical system where a periodic signal can be detected with the PDM or ACF. Alternately, for a given input set of parameters for a periodically-induced signal, one can estimate the RMS and then power ratio of the signal against the red noise continuum, and compare to threshold values of log($P_{\rm rat}$) that can be detected with the PDM or ACF. \textit{Moreover, given the high minimum values of log($P_{\rm rat}$) needed for detection as demonstrated by this paper, when one claims detection of a periodic signal using the ACF or PDM, they are implicitly claiming a minimum value for log($P_{\rm rat}$) and thus RMS --- and these values have implications for the physical parmaeters of a given system.}

Here, we briefly discuss the application to periodically self-lensing SMBH binaries in highly-inclined systems on using the results from ACF \& PDM.

\subsubsection{Highly-inclined self-lensing SMBH binaries}

We consider highly-inclined super massive black hole (SMBH) binary systems following the framework of \citet{2018MNRAS.474.2975D}; hereafter \hyperlink{MyKeyH}{DD18}. Emission from the accretion disk around the primary black hole is periodically gravitationally lensed and magnified by the secondary. For simplicity, we assume that our simulations results, derived assuming QPOs with a very high quality factor, are applicable to such strictly-periodic systems. SMBH binary systems could additionally have quasi-periodic emission, potentially. Hydrodynamic simulations of accretion onto SMBH binaries from circumbinary disks can yield quasi-periodic fluctuations in accretion rate through the circumbinary disk (e.g. \citealt{2014ApJ...783..134F}; \citealt{2015MNRAS.446L..36F}). If such variations in accretion rate directly yield variations in luminosity when that material accretes onto the black holes, then quasi-periodic components contributing to observed emission might be expected.

However, tests incorporating mixtures of QPOs and red noise (including that generated in the two accretion mini-disks) would be difficult, because predicting the degree and PSD shape of the contaminating red noise is quite challenging: there could be additional modes of variability not present in stable/ more persistently-accreting disks; one would need to know the radii over which those variability mechanisms are triggered, taking into account e.g. tidal truncation of the outer disks (e.g. \citealt{1977ApJ...216..822P}; \citealt{2016ApJ...828...68N}). Consequently consideration of red noise (from mini-disks) with QPO mixtures in this context is beyond the scope of our paper.

We consider just the optical and UV emission from the inner accretion disk. We use \citet{soton161201}, their Fig.\ 5.3, as an empirical guide for optical PSD slopes and normalization. \citet{soton161201} publish V-band unbroken power-law PSDs for seven Seyferts; we take the average value of slope ($\beta=2.3$) and normalization (20 Hz$^{-1}$ at $\rm 10^{-6}$ Hz).

We note that any possible evolution with black hole mass ($M_{BH}$) or accretion rate relative to Eddington $L_{\rm Bol}/L_{\rm Edd}$ (as is known for broadband X-ray PSDs of nearby Seyferts, e.g. \citealt{2006Natur.444..730M}) is not accounted for here. Any given SMBH binary system may very likely host black holes with differing values of black hole masses and if both the black holes in the binary system are accreting, then $L_{\rm Bol}/L_{\rm Edd}$ might also differ and yield different individual PSD shapes and/or normalizations, and the sum of their variability characteristics will be reflected on the single optical PSD shape that we measure. Hence, for extreme simplicity, we neglect such effects and assume that the single PSD shape represents the level of red noise emission from both systems.

We adopt from DD18 magnification factors for binary systems having mass ratios $q=0.05$ and $q=0.5$, and having the accretion disks that are nearly edge-on, with inclination $J=0.2$ rad with respect to the line of sight. We consider each of the NUV and V bands separately; the former represents a more ``optimistic'' case in terms of the magnification. We adopt the waveforms in Fig.~3 of DD18 to estimate the RMS/mean for each of four binary orbital inclinations to the line of sight in units of the number of Einstein radii, $N_E$, which quantifies the angular separation of source and lens at closest approach, corresponding to $N_E$ = 0.05, 0.1, 0.5, and 1.0. A magnification factor is not the same as a change in flux, since the input flux from the accretion disk, which is stochastic, will vary with time. For simplicity, though, we ignore this effect. Values of RMS/mean are listed in Table~\ref{tab:rmswithlogpr}.

We simulate 250-day long continuous monitoring light curves that are evenly sampled, assuming a $Q=30$ QPO at the MF frequency used in Sections 3--5, $8\times10^{-7}$ Hz, and considering each of the values of RMS/mean for each inclination/waveband/mass ratio combination, mixed with a $\beta=2.3$ continuum PSD as per the parameters listed above. For simplicity, we assume the same continuum PSD parameters for both the V and NUV bands. The resulting values of log($P_{\rm rat}$) are listed in Table~\ref{tab:rmswithlogpr}; they span from roughly 2.9 for the $N_{\rm E}=1.0$ cases up to 5.7 for the $N_{\rm E}=0.05$ cases.

Recall, however, from $\S$4, Fig.~\ref{fig:QRNPDM} that when using the PDM, for $\beta=2.3$, values of log($P_{\rm rat}$) of 5 were required for robust true positive detections at $\sim$ 99 per cent confidence for the MF QPO. For the ACF, values of log($P_{\rm rat}$) of 5 were required for robust true positive detections at $\gtrsim$ 99.7 per cent confidence; the $N_{\rm E}=0.05$ and 0.1 cases would thus likely be detected. (Note: we choose to be approximate in our conclusions here given the large uncertainties and assumptions regarding the form of the continuum PSD as discussed above.)

Note that we are commenting on the possibility of detections in individual systems only; determining the frequency of occurrence of such systems whose orbital parameters satisfy these requirements is beyond the scope of this paper.

\section{CONCLUSIONS}
In this paper, we have presented results from an empirical investigation of the behavior of two statistical tools --- the ACF and the PDM --- when used for detecting a QPO in light curves that contain broadband stochastic red noise variability.

These tools have been used to claim the presence of periodic signals in AGN light curves; however, pure red-noise processes can easily mimic features that could falsely be intepreted as a QPO. Guidance in both preventing false-positive QPO claims and in supporting true-positive detections is needed, particularly given that the astronomical community now has, or will have access to databases containing large numbers of monitoring light curves courtesy of current and near-future large-area ground-based monitoring programmes such as LSST.
  
This paper is intended to provide guidance both to those examining individual light curves for QPOs and to those performing data trawls in large databases. Our overarching goal is to help reduce the appearance of false claims of periodocities (features consistent with the null hypothesis of simple red noise) in the literature.

We perform Monte Carlo simulations covering a range of red noise power-law PSD slopes and QPO strengths for a few select light curve sampling patterns. We empirically investigate both true-positive detections of QPOs when mixtures of broadband red noise and a narrow-band QPO are present and false positive detections when no QPO is present, only red noise (the null hypothesis model).

We determine that pure red noise light curves tested for a broad range of power-law slopes were able to produce false positive signals in the ACF more than $0.3$ per cent of the time at all slopes tested, considering the full timescale range.  False-positive peaks routinely reach correlation coefficients of 0.5 and higher. When we restrict ourselves to lags less than one-third of the full duration, the rate of false positives still remains $\geq 0.3$ per cent at power-law slopes $\beta \gtrsim 2.6$, though maximum correlation coefficients remain below 0.55. We conclude that if one observes a peak with correlation coefficient $>$0.5 and while restricting oneself to lags less than roughly 1/3 of the duration, and given even data sampling, then the signal is probably real.

In the case of the PDM, pure red-noise processes cause a dips in the value of the PDM test statistic $\theta$ (e.g., $\theta < \sim 0.6$), particulary at frequencies corresponding to timescales greater than roughly one third of the duration. The rate of false positives is greater than $0.3$ per cent across all the tested slopes, and at much steeper slopes $\beta \gtrsim 1.4$, false positives occur almost 99.9 per cent of the time. However, when we simply neglect the lowest frequencies (timescales greater than roughly one third of the duration), the rate of false positives is still greater than $0.3$ per cent for all the slopes, but having the PDM statistic $\theta$ below 0.65 is $<0.3$ per cent. Hence, neglecting the lower frequency bins in the PDM is a straightforward way to minimize false postitive detections.

These results are an empirical demonstration that, when using the ACF or PDM, and when variability due to a pure red noise process with an unbroken power-law PSD shape, the ad-hoc action of simply disregarding timescales greater than $\sim$1/3 of the full duration is an effective way to reduce false positives (we strongly caution, however, that this one-third benchmark is meant to be highly approximate only, and can depend on
e.g., steepness of power-law slope, data sampling, etc.). They also suggest that features occurring at the longest timescales/lowest frequencies in previously-published ACFs/PDMs may have been consistent with pure red noise instead of being due to QPOs, as claimed.

When there are QPOs mixed with pure red noise in evenly-sampled data, the true-positive detection sensitivity in both the ACF and PDM naturally depends strongly on not just the relative strengths of the QPO and the red noise (log($P_{\rm rat}$)) but also sensitively on the steepness of the red noise PSD slope.  We find that extremely large values of log($P_{\rm rat}$) typically 4--5 without the pre-filtering technique introduced in $\S$3.3 \& $\S$4.3 --- are typically required for a 99.7 per cent true-positive detection rate, at PSD slope --- $\beta$ typically $<$ $\sim$2.

The reader is reminded that we did not add Poisson noise to our simulated light curves; this would likely make a true-positive detection even more difficult, so our results are effectively lower limits to values of log($P_{\rm rat}$).

We re-iterate that any claim of a QPO detection using the ACF or PDM implies detection of a signal with such a high power ratio. Depending on the model applied for intepretation, it is possible that the high values of log($P_{\rm rat}$) and RMS imply extreme regions of model parameter space. For example, in the specific case of periodically self-lensing black holes in gravitationally-closed binary systems on a highly-inclined orbit that we considered in $\S$6.2, values of log($P_{\rm rat}$) $\gtrsim 5$ (assuming no light curve filtering) in optical/UV emission from the inner disk require the source and lens to be separated by only $N_{E} \sim$ 0.05 at closest approach (DD18).

We also examined a small range of different sampling patterns, including yearly sun gaps as is common to ground-based optical observing programs, and irregular but sustained sampling, and their effect on true- and false-positive detection rates in the PDM; True-positive detection rates for a given values of log($P_{\rm rat}$) and $\beta$ are not significantly impacted.

If readers who are searching for QPOs have light curves with sampling patterns identical to those we simulated, then we encourage those readers to use our simulations and plots as guides to determining whether features in their own ACFs/PDMs are indicative of any QPO or are consistent with a single pure red noise process.

If readers have light curves with different sampling patterns, and/or want to test a null hypothesis model more complex than a simple unbroken power-law, then we recommend that they perform their own simulations, including testing the range of behavior that can be exhibited under the appropriate null hypothesis model, depending on the slope of the underlying red noise process (and its uncertainty). \textit{If the slope is not known specifically, then readers should test a wide range of slopes in order to be conservative.} As a final reminder, we did not consider the effects of Poisson noise in our simulations here, and readers should take that into account in their own simulations.

%-------------------

It is worth considering whether or not a periodogram is more efficient at cleanly separating broadband continuum noise from a narrow-band QPO compared to the ACF or PDM for a given sampling pattern.
For evenly-sampling data, the method of \citet{2005A&A...431..391V} is applicable for QPO detection in a periodogram, and the behavior of the periodogram (probability distribution of powers, biases, the extent to which adjacent temporal frequencies are independent, etc.) has been well understood (see references in e.g. \citealt{2005A&A...431..391V}).
However, for a given RMS strength, the value of log($P_{\rm rat}$) needed to register a detection of a QPO will depend on factors such as temporal frequency spacing and the quality factor $Q$. For unevenly-sampled data, other factors come into play for all methods, such as the impact of aliasing at high temporal frequencies, which depends on spectral slope and data sampling. Such detailed discussions are beyond the scope and intent of the present paper, but we encourge readers who have a specific sampling pattern in mind to perform simulations for all methods to help gauge detection rates of both true and false positives. We would like to warn users of the ACF and PDM that it is not possible to use these tools to reliably seperate a narrow-band signal and the red noise continuum and quantify the form of that continuum, since the presence of red noise causes the ACF and PDM values to become highly self-correlated.

We would thus encourage the community to consider testing of additional methods, including Bayesian fits (e.g. CARMA; \citealt{2020ApJ...900..117Z}), and development of automatic light curve classifiers (e.g. \citealt{2021AJ....161..141S}) that could potentially quantify deviations from pure-red noise behavior. However, comparison testing of how the various methods --- be they classical such as ACF, PDM, periodogram, or modern --- perform in maximizing true positive detections and minimizing false detections is beyond the scope of the current paper.

%-------------------

Finally, we strongly encourage readers to report the implied value of the RMS (or equivalently, RMS/mean) of the QPO, in addition to timescale, in any future QPO claims they publish, regardless of the statistical method used. The RMS encodes additional information about the physical parameters of the periodic process and/or variability mechanism (e.g., observer viewing angles; jet angle parameters; energy dependence of a variability mechanism), yet such information is regularly neglected in published QPO claims.
Hopefully, physical parameters implied by the QPO strength will be consistent with physical information obtained from other avenues methods (imaging, spectroscopy), if available.

\section*{ACKNOWLEDGEMENTS}
SK, AM, and ASC acknowledge support from Narodowe Centrum Nauki (NCN) grant 2016/23/B/ST9/03123.
SK and AM also acknowledge partial support from NCN grant 2018/31/G/ST9/03224.
SK and AM thank T.\ Bogdanovic for useful discussions on applications
to physical systems.

\section*{DATA AVAILABILITY}
The data underlying this article will be shared on reasonable request
to the corresponding author.

\section*{APPENDIX}

\begin{table*}
\begin{tabular}{llll}
\hline
\hline
\multicolumn{1}{c}{} & \multicolumn{3}{c}{PARAMETER SPACE for significant detection on using the ACF }                                                                                                                           \\ \hline
\multicolumn{1}{c}{Centroid Frequency} & \multicolumn{1}{c}{TRUE POSITIVES}                                    & \multicolumn{1}{c}{TYPE II ERROR }                             & \multicolumn{1}{c}{ERROR 3 }                                    \\ 
\multicolumn{1}{c}{$f_L$ ($10^{-7}$ Hz)} & \multicolumn{1}{c}{ ($99.7\%$) }                                    & \multicolumn{1}{c}{($\lesssim 0.3 \%$)  }                             & \multicolumn{1}{c}{($\lesssim 0.3 \%$)  }  \\ \hline
$ 8.0$ (MF QPO)                  & \multicolumn{1}{l}{log($ P_{\rm rat}) \sim 5/4$ at $\beta \lesssim 2.4/1.8$}      & \multicolumn{1}{l}{log($ P_{\rm rat}) \sim 5/4$ at $\beta \lesssim 2.4/1.8$}      & \multicolumn{1}{l}{log($ P_{\rm rat}) \sim 5/4$ at $\beta \lesssim 3.0/1.8$}      \\ \hline
$32.0$ (HF QPO)                 & \multicolumn{1}{l}{log($P_{\rm rat}) \sim 5/4/3$ at $\beta \lesssim 2.2/1.6/1.2$} & \multicolumn{1}{l}{log($P_{\rm rat}) \sim 5/4/3$ at $\beta \lesssim 2.2/1.6/1.2$} & \multicolumn{1}{l}{log($P_{\rm rat}) \sim 5/4/3$ at $\beta \lesssim 2.4/1.6/1.2$} \\ \hline
\hline
\end{tabular}
\caption{The parameter space for significant detection of the period of the MF \& HF QPO signals mixed with red noise of unbroken PL model on using the ACF  tested for evenly sampled light curves in $\S$3.}
\end{table*}

\begin{table*}
\begin{tabular}{llll}
\hline
\hline
\multicolumn{1}{c}{} & \multicolumn{3}{c}{PARAMETER SPACE for significant detection on using the PDM }                                                                                                                           \\ \hline
\multicolumn{1}{c}{Centroid Frequency} & \multicolumn{1}{c}{TRUE POSITIVES}                                    & \multicolumn{1}{c}{TYPE II ERROR }                             & \multicolumn{1}{c}{ERROR 3 }                                    \\ 
\multicolumn{1}{c}{$f_L$ ($10^{-7}$ Hz)} & \multicolumn{1}{c}{}                                    & \multicolumn{1}{c}{ }                             & \multicolumn{1}{l}{ }  \\ \hline
$ 8.0$ (MF QPO)                  & \multicolumn{1}{c}{log($ P_{\rm rat}) \sim 5/4$ at $\beta \lesssim 2.4/1.8$} ($99\%$)     & \multicolumn{1}{c}{log($ P_{\rm rat}) \sim 5/4$ at $\beta \lesssim 2.4/2.0$}  ($\lesssim 0.3 \%$)    & \multicolumn{1}{c}{log($ P_{\rm rat}) \sim 5/4$  at $\beta \lesssim 2.6$} ($\lesssim 1\%$)     \\ \hline
$32.0$ (HF QPO)                  & \multicolumn{1}{c}{log($P_{\rm rat}) \sim 5/4/3$ at $\beta \lesssim 2.0/1.6/1.0$} ($99.7\%$)   & \multicolumn{1}{c}{log($P_{\rm rat}) \sim 5/4/3$ at $\beta \lesssim 2.0/1.6/1.0$} ($\lesssim 0.3 \%$) & \multicolumn{1}{c}{log($P_{\rm rat}) \sim 5/4$ at $\beta \lesssim 2.6$}($\lesssim 1\%$) \\ \hline
\hline
\end{tabular}
\caption{The parameter space for significant detection of the period of the MF \& HF QPO signals mixed with red noise of unbroken PL model on using the PDM  tested for evenly sampled light curves in $\S$4.}
\end{table*}
\bibliographystyle{mnras}
\bibliography{references} % if your bibtex file is called example.bib

%\section*{APPENDIX}
%\input{appendixv2.tex}
%\begin{thebibliography}{99}
%\end{thebibliography}

%%%%%%%%%%%%%%%%%%%%%%%%%%%%%%%%%%%%%%%%%%%%%%%%%%

%%%%%%%%%%%%%%%%% APPENDICES %%%%%%%%%%%%%%%%%%%%%
\iffalse
\appendix

\section{Some extra material}

If you want to present additional material which would interrupt the flow of the main paper,
it can be placed in an Appendix which appears after the list of references.

%\input{appendix.tex}

%%%%%%%%%%%%%%%%%%%%%%%%%%%%%%%%%%%%%%%%%%%%%%%%%%

\fi
% Don't change these lines
\bsp	% typesetting comment
\label{lastpage}

\end{document}